%% file: HybridLunarPNT_v1.tex
\documentclass[letterpaper]{article}
\usepackage[left=.8in,right=.7in,top=1in,bottom=1in]{geometry}

\include{mydefs_arxiv.tex}

\begin{document}
\begin{acronym}
	\input{acronyms}
\end{acronym}

\title{Towards Hybrid Lunar PNT: Error Models, Lower Bounds and Algorithms}
\author{\normalsize Robert P\"ohlmann, Emanuel Staudinger\\
		\normalsize Institute of Communications and Navigation\\
		\normalsize German Aerospace Center (DLR)\\
		\normalsize Münchener Str. 20\\ 
		\normalsize 82234 Wessling,
		\normalsize Germany\\
		\normalsize robert.poehlmann@dlr.de
		\and
		\normalsize Gonzalo Seco-Granados\\		
		\normalsize Dept. of Telecommunications and Systems Engineering\\
		\normalsize Center for Space Studies and Research (CERES), IEEC\\
		\normalsize Universitat Autònoma de Barcelona (UAB)\\
		\normalsize 08193 Bellaterra,
		\normalsize Spain\\
		\normalsize gonzalo.seco@uab.cat}
\date{}
\maketitle

\begin{abstract}
Accurate \ac{PNT} are crucial for upcoming lunar surface missions. Lunar satellite navigation systems are being developed, but lack coverage during early deployment phases.
Hybrid lunar \ac{PNT} combining cooperative navigation, satellite systems, and an optional reference station offers improved accuracy and availability.
This study develops realistic error models that incorporate temporal correlations often ignored in existing works. We derive a cooperative navigation error model considering fading and pseudorange bias from multipath propagation, and compare three error models for lunar satellite pseudorange and pseudorange rate signal-in-space error.
These temporal error correlation models integrate easily into Kalman filters and provide realistic performance predictions essential for robust navigation engines.
We perform case studies to demonstrate that hybrid navigation significantly improves accuracy, particularly with static users present. Most notably, hybrid navigation enables optimal performance when using a lunar reference station, achieving sub-meter accuracy with only two visible satellites.
\end{abstract}

\acresetall
\section{Introduction}
According to the \citet{internationalspaceexplorationcoordinationgroup2022}, a large number of lunar surface missions are planned in the coming years. Mission objectives include science, sustained human presence on the Moon, mining of natural resources, and the preparation of human missions to Mars.
Accurate and reliable \ac{PNT} are key requirements, e.g. for autonomous robotic exploration, pinpointing scientific measurements and localization of astronauts.
\citet{wallace2024} have analyzed required position accuracies, reporting meter level for human exploration, decimeter level for robotic mining and centimeter level for certain scientific measurements.

Past missions mostly relied on tracking from Earth, e.g. by the Universal Space Network or the Deep Space Network.
The use of ground stations is costly and does not scale for a growing number of missions.
Hence, using weak \ac{GNSS} signals from Earth for lunar positioning is an active research topic \citep{manzano2014,musumeci2016,iiyama2024a}.
The \ac{LuGRE} mission aims to demonstrate \ac{GNSS} signal reception and navigation in lunar orbit and on the surface \citep{konitzer2024}.
Both tracking from Earth and receiving \ac{GNSS} signals are limited to spacecrafts on the nearside of the Moon.
To achieve global coverage, dedicated lunar satellite communications and navigation systems have been suggested;
JAXA's \ac{LNSS} \citep{murata2022a},
NASA' \ac{LCRNS} \citep{nasa2022}, and
ESA's \ac{LCNS} \citep{grenier2022}, developed as part of ESA's Moonlight initiative.
All of these systems take heritage from \ac{GNSS}.
Following interoperability and standardization efforts under the frame of LunaNet \citep{nasa2025}, users shall be able to jointly use signals received from different systems. The \ac{AFS} has been standardized \citep{nasa2025a}. 
Especially in an early phase, there will only be a small number of LunaNet satellites with navigation capabilities available. Thus, availability and accuracy will be limited, even more so in areas with obstructed satellite visibility like craters and skylights.
Furthermore, \ac{PNT} capabilities shall not solely be developed for the Moon, but shall later be extended to Mars following the Moon to Mars concept \citep{nasa2024,internationalspaceexplorationcoordinationgroup2022}. For Mars, the number of navigation satellites will likely be even more limited.
In order to overcome these limitations, \citet{melman2022,audet2024} have investigated fusion of \ac{LCNS} with a \ac{DEM} for robotic surface missions.
\citet{wallace2024} have proposed pseudolites to illuminate a crater with navigation signals.
Furthermore, a lunar reference station could increase \ac{PNT} availability and accuracy and
is considered as low-cost and fast way forward by \citet{jun2025}, who also present a concept of operations and analyze available hardware  technology readiness levels.
A lunar reference station could provide higher navigation accuracy by differential corrections \citep{psychas2024,jun2025} and increased communications data rate \citep{jun2025}.

A broader concept is hybrid navigation, combining lunar satellite and cooperative surface navigation, suggested by \citet{pohlmann2024}. It can include a lunar reference station.
Hybrid navigation builds on the complementary nature of satellite and cooperative navigation. Satellite navigation enables \ac{PNT} with respect to a global reference frame and time, subject to the number of visible satellites. Cooperative navigation offers high accuracy and availability in the areas where users are present.
The hybrid approach is not limited to a specific lunar surface wireless communications system. Any system that supports one-way \ac{ToF} measurements is in general suitable, e.g. the swarm navigation system by \ac{DLR} \citep{zhang2021,pohlmann2023}, IEEE 802.11 (WiFi) or 3GPP (4G/5G) mentioned by \ac{LNIS} \citep{nasa2024}, as well as other technologies like \ac{UWB}. For improved performance, satellite and cooperative navigation subsystems shall use the same oscillator, which is called physical layer cooperation by \citet{pohlmann2024}. 
Many papers on lunar \ac{PNT} assume white Gaussian noise for the pseudorange \ac{SISE} and thereby ignore temporal error correlations, e.g.
\citet{grenier2022,melman2022,pohlmann2024}.
In contrast, \citet{audet2024} include a sophisticated simulation of lunar \ac{SISE} in their analysis, but the simulator is not open source.
Regarding lunar surface pseudoranging, \citet{wallace2024} do not consider spatial or temporal correlation of multipath induced errors.
\citet{jun2025} investigate multipath effects on lunar surface communications using raytracing and the two-ray ground reflection model to calculate path loss. However, bias effects on (pseudo)-ranging are not regarded.
Error models for lunar satellite and cooperative surface navigation, which consider the temporal correlation of pseudorange and pseudorange rate errors, have not been published yet. Open source parametric error models make results reproducible by the community. Furthermore, such error models can be applied in an augmented navigation filter for improved performance and robustness.

With this paper, we provide the following key contributions:
\begin{itemize}
	\item We derive an error model for cooperative radio navigation on the lunar surface, which takes fading and pseudorange bias due to multipath propagation into account.
	\item We present and compare three different error models for the lunar satellite navigation \ac{SISE} of pseudorange and pseudorange rate.
	\item We perform three case studies of hybrid lunar \ac{PNT}, including different operation modes of a lunar reference station, by investigating lower bounds on the estimation error.
	\item We compare the performance of three Kalman filter variants for hybrid lunar \ac{PNT}, which use an augmented state space with temporally correlated error models.
\end{itemize}

The remainder of this paper is organized as follows. \Cref{s:coopNav,s:satNav} derive the cooperative and satellite navigation error models, respectively.
\Cref{s:hybrid} introduces the hybrid navigation system model including the augmented state space.
\Cref{s:bcrb} presents the \ac{BCRB} for tracking.
\Cref{s:algorithms} defines three Kalman filter based algorithms for hybrid lunar \ac{PNT}, operating on the augmented state space. 
\Cref{s:case} conducts three case studies of hybrid lunar \ac{PNT} based on the \ac{BCRB}.
\Cref{s:results} presents simulation results of the hybrid navigation algorithms.
\Cref{s:conclusion} concludes the paper.

\section{Cooperative Navigation Error Model}\label{s:coopNav}
\subsection{Two-Ray Ground Reflection Model}
\begin{figure}[htb]
	\begin{minipage}{0.6\textwidth}
		\centering
		\includegraphics[width=\textwidth]{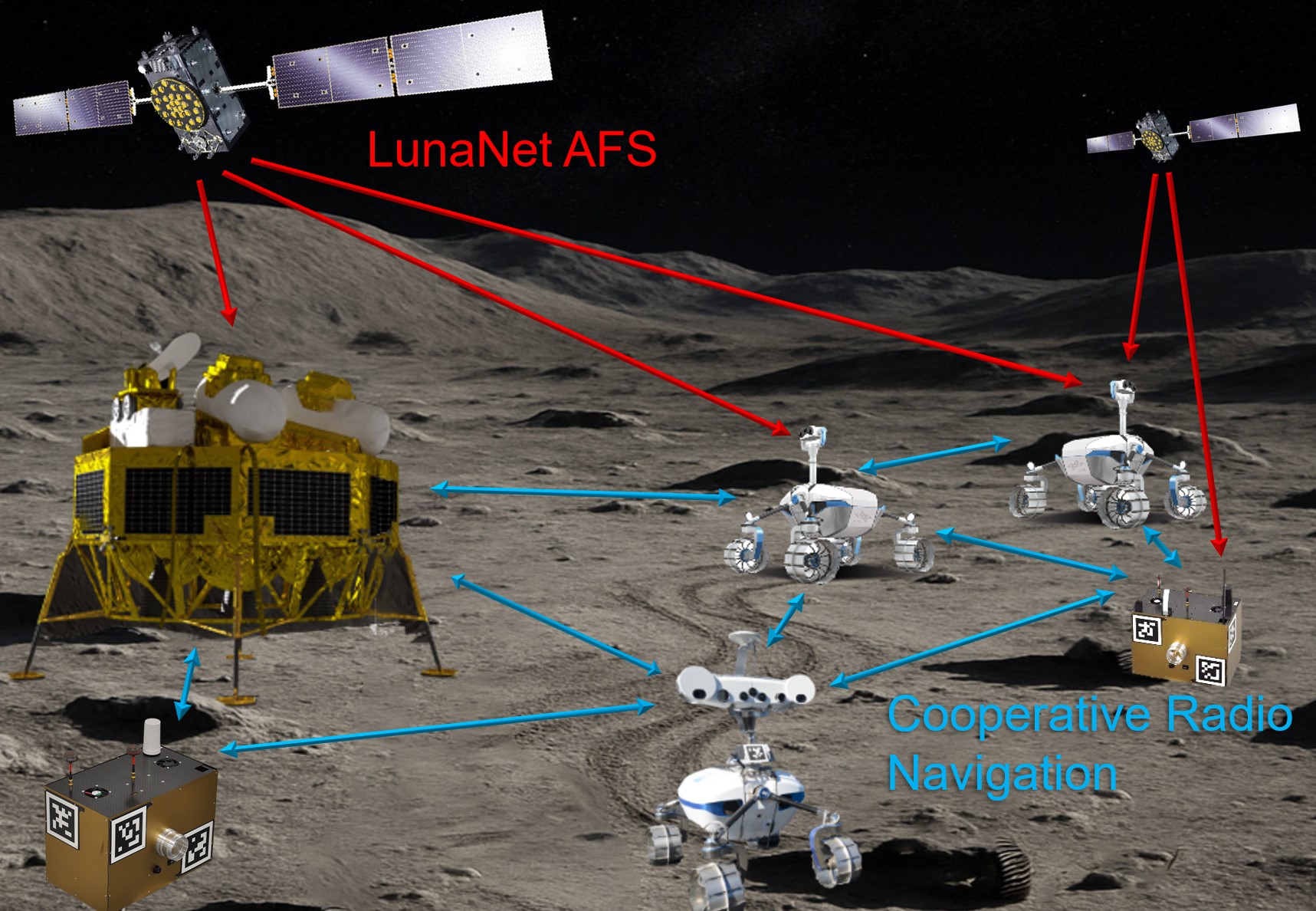}
		\caption{Lunar hybrid navigation scenario with satellites transmitting LunaNet AFS signals and cooperative radio navigation among entities on the lunar surface}
		\label{fig:LunarHybridNavigation}
	\end{minipage}
	\hfill
	\begin{minipage}{0.37\textwidth}
		\centering
		\centering
		\includegraphics{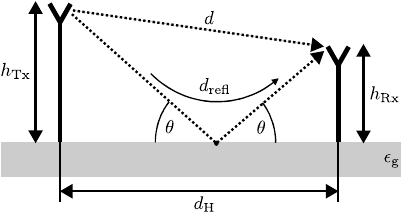}
		\caption{Two-ray ground reflection model}
		\label{fig:2rayModel}
	\end{minipage}
\end{figure}

In this paper, we investigate hybrid lunar navigation with lunar satellites transmitting LunaNet \ac{AFS} signals and a lander, robotic rovers and instrument packages exchanging signals for cooperative radio navigation, see \cref{fig:LunarHybridNavigation}. In this section, we start with the cooperative part.
For a clear distinction, we refer to signals exchanged among nodes on the lunar surface as "cooperative", although the following link-level analysis is independent of the actual cooperation, which happens on the localization layer.

To model radio propagation on the lunar surface, we use the two-ray ground reflection channel, which has been widely used to model large-scale fading in narrowband wireless communications \citep{rappaport2002}. Its impact on cooperative ranging and navigation has been investigated by \citet{staudinger2023}.
\Cref{fig:2rayModel} shows the typical setting where transmitter and receiver are located at height $h_\mathrm{Tx}$ and $h_\mathrm{Rx}$, respectively, and are separated by the horizontal distance $h_\mathrm{H}$.
The signal travels from transmitter to receiver via two rays.
The first ray is the \ac{LoS} component with distance $d$.
The second ray is the ground reflected component with equivalent distance $d_{\mathrm{refl}} = \sqrt{(h_{\mathrm{Tx}} + h_{\mathrm{Rx}})^2 + d_{\mathrm{H}}^2}$ and incident angle $\theta$.
Electrical properties of the ground are covered by its complex permittivity $\epsilon_{\mathrm{g}}$.
The real part of $\epsilon_{\mathrm{g}}$ is the relative permittivity, the imaginary part is the conductivity.
At the receiver, both signals are superposed, thus the received signal as a function of time $t$ is given by
\begin{equation}\label{eq:rsig}
	r\left(t\right) = r_\text{LoS}\left(t\right) + r_\text{refl}\left(t\right)
	= \frac{\lambda_{\mathrm{c}}}{2\pi d}s\left(t-\frac{d}{c}\right)\text{e}^{-\cj\frac{2\pi d}{\lambda_{\mathrm{c}}}} + \Gamma\left(\theta\right)\frac{\lambda_{\mathrm{c}}}{2\pi d_{\text{refl}}}s\left(t-\frac{d_{\text{refl}}}{c}\right)\text{e}^{-\cj\frac{2\pi d_{\text{refl}}}{\lambda_{\mathrm{c}}}},
\end{equation}
where $c$ is the speed of light and $\lambda_{\mathrm{c}} = c / f_{\mathrm{c}}$ is the carrier wavelength.
The transmitted signal $s(t)$ is delayed by $d/c$ and
the reflected signal is delayed by $\frac{d_{\text{refl}}}{c}$. %
The reflection coefficient for circular co-polarization $\Gamma(\theta)$ is calculated based on the respective reflection coefficients for linear vertical and horizontal polarization following \citet{hannah2001},
\begin{equation}
	\Gamma(\theta) = \frac{\Gamma_\mathrm{v}(\theta) + \Gamma_\mathrm{h}(\theta)}{2}, \qquad
	\Gamma_{\mathrm{v}}(\theta) = \frac{\epsilon_{\mathrm{g}} \sin(\theta) - \sqrt{\epsilon_{\mathrm{g}} - \cos^2(\theta)}}{\epsilon_{\mathrm{g}} \sin(\theta) + \sqrt{\epsilon_{\mathrm{g}} - \cos^2(\theta)}}, \qquad
	\Gamma_{\mathrm{h}}(\theta) = \frac{\sin(\theta) - \sqrt{\epsilon_{\mathrm{g}} - \cos^2(\theta)}}{\sin(\theta) + \sqrt{\epsilon_{\mathrm{g}} - \cos^2(\theta)}}.
\end{equation}
From \cref{eq:rsig}, we determine the received signal power of a narrowband signal according to the two-ray ground reflection model as
\begin{equation}\label{eq:Prx}
	P_{\mathrm{Rx}}\left(d_\mathrm{H}\right) = P_{\mathrm{Tx}}\left(\frac{\lambda_{\mathrm{c}}}{2\pi}\right)^2 \left\lvert \frac{1}{d} + \Gamma\left(\theta\right)\frac{1}{d_{\text{refl}}}\text{e}^{-\cj\Delta\phi} \right\rvert^2, \qquad
	\Delta\phi = \frac{2\pi}{\lambda}\left(d_{\text{refl}}-d\right).
\end{equation}
Both, the direct and the reflected component experience free-space path loss. At the receiver, they superpose constructively or destructively, depending on the carrier phase difference.
Power and phase difference of the reflected component with respect to \ac{LoS} are determined by
the complex permittivity of the ground $\epsilon_{\mathrm{g}}$,
the distance difference $d_{\text{refl}}-d$
and the incident angle $\theta$.

\subsection{One-Way Time-of-Flight Radio Ranging}\label{ss:tof}
We consider that one-way \ac{ToF} radio ranging is performed among all users on the lunar surface which are part of the set $\mathds{U}$. A user can e.g. be a robot, an astronaut, or a static instrument package.
The pseudorange for a signal transmitted by user $j$ and received by user $i$ at epoch $k$ is given by
\begin{equation}\label{eq:prCoop}
	\rho_{i,j}^k = \norm{\bm{p}_j^k - \bm{p}_i^k} + c\delta_i^k - c\delta_j^k + \epsilon_{i,j}^{\rho,k},\: i,j \in \mathds{U},
\end{equation}
where $\bm{p}_i^k$ is the three-dimensional user position. The clock offset $\delta_i^k$ is multiplied by the speed of light $c$. The quantities for user $j$ are defined analogously.
We model the cooperative pseudorange error as
\begin{equation}\label{eq:prCoopError}
	\epsilon_{i,j}^{\rho,k} \sim \mathcal{N}\left(b_{\mathrm{coop},l(i,j)}^k,\, (\sigma_{\mathrm{ToF}}^k)^2\right),
\end{equation}
with a time-varying bias due to the two-ray ground reflection $b_{\mathrm{coop},l(i,j)}^k$ for link index $l(i,j)$,
and Gaussian measurement noise with variance $(\sigma_{\mathrm{ToF}}^k)^2$ depending on the \ac{SNR}.
In the following, we derive the cooperative pseudorange variance $(\sigma_{\mathrm{ToF}}^k)^2$.
In the next \cref{ss:coopGMP}, we model the cooperative pseudorange bias $b_{\mathrm{coop},l}^k$ as \ac{GMP-1} based on the two-ray ground reflection model.
For notational clarity, we drop the subscripts where they are not needed.

The variance of the estimated pseudorange $\hat{\rho}$ is lower bounded by the \ac{CRB} for \ac{ToF} ranging,
\begin{equation}\label{eq:tofCRB}
	\text{var}\left\lbrace \hat{\rho} \right\rbrace \geq \operatorname{CRB}(\hat{\rho}) = \frac{c^2}{8 \pi^2 \frac{E_{\mathrm{s}}}{N_0} \bar{\beta}^2}, \qquad
	\bar{\beta}^2 = \frac{\int f^2\lvert S(f)\rvert^2\text{d}f}{\int \lvert S(f)\rvert^2\text{d}f},
\end{equation}
with the symbol energy to noise power spectral density ratio $\frac{E_{\mathrm{s}}}{N_0}$ and  the mean square bandwidth  of the signal $\bar{\beta}^2$ \citep{dardari2009}.

\Ac{OFDM} is the basis for many state-of-the-art wireless communications systems like WiFi and 4G/5G cellular networks. It is also considered for lunar surface communications \citep{nasa2025} and applied in \ac{DLR}'s swarm navigation system introduced in the next section. An \ac{OFDM} signal in baseband can be written as
\begin{equation}
	s(t) = \frac{1}{\sqrt{N_{\mathrm{fft}}}} \sum\limits_{n=-N_{\mathrm{fft}}/2}^{N_{\mathrm{fft}}/2-1} S(n) \, e^{\cj 2 \pi n f_{\mathrm{sc}} t},
\end{equation}
with subcarrier index $n$,
complex symbol $S(n)$,
sampling rate $B$ and subcarrier spacing $f_{\mathrm{sc}} = \frac{B}{N_{\mathrm{fft}}}$ with \ac{FFT} length $N_{\mathrm{fft}}$.
Estimation variance is only one error source. Another error source is the estimation bias, which depends on the specific estimator. We use a \ac{ML} estimator in frequency domain
\begin{equation}\label{eq:ml}
	\hat{\rho} = \arg\max_{\rho} \left\lvert\sum\limits_{n=-N_{\mathrm{fft}}/2}^{N_{\mathrm{fft}}/2-1} R(n) S^*(n) e^{\cj 2 \pi n f_{\mathrm{sc}} \rho/c}\right\rvert,
\end{equation}
where $R(n)$ are frequency domain samples of the received signal, obtained by sampling $r(t)$ from \cref{eq:rsig} and applying \ac{FFT}.
Since $r(t)$ contains not only the \ac{LoS} component, but also a closely spaced multipath component from the ground reflection, the pseudorange estimate $\hat{\rho}$ is biased. 
Its estimation bias $B(\hat{\rho}) = \E{\hat{\rho}} - \rho$, as well as the derivative $\nabla_\rho B(\hat{\rho})$, can be determined by simulation based on the two-ray ground reflection model \citep{staudinger2023}.
Ultimately, we are interested in the conditional estimation \ac{MSE}, which considers both estimation variance and bias and following \citet{vantrees2007} is lower bounded by
\begin{equation}\label{eq:MSE}
	\operatorname{MSE}\{\hat{\rho}\} = \E{(\hat{\rho}-\rho)^2} \geq \operatorname{CRB}(\hat{\rho}) + \operatorname{B}^2(\hat{\rho}) + \operatorname{CRB}(\hat{\rho}) \left( 2 \nabla_\rho B(\hat{\rho}) + \nabla_\rho B^2(\hat{\rho}) \right).
\end{equation}

\subsection{Cooperative Pseudorange Error}\label{ss:coopPRError}
The actual cooperative navigation error model parameters depend on the communications system parameters.
We consider \ac{DLR}'s swarm navigation system for local radio communications and navigation on the lunar surface, which enables radio communication and navigation for planetary exploration \citep{zhang2021,pohlmann2023}. Decentralized estimation algorithms require frequent exchange of moderate-sized data packets \citep{pohlmann2025a}, while downloading scientific data or teleoperating robots demand high data rates. To address these needs, the system design prioritizes bandwidth, update rate and ranging performance.
The communications system is based on \ac{OFDM} with
carrier frequency $f_\mathrm{c} = \unit[2]{GHz}$,
signal bandwidth $B = \unit[10]{MHz}$,
\ac{OFDM} with symbol length $N_{\mathrm{fft}} = 1024$ and 922 allocated subcarriers,
transmit power $P_{\mathrm{Tx}} = \unit[100]{mW}$, %
receiver temperature $\unit[290]{K}$ and receiver noise figure $\unit[5]{dB}$,
\ac{RHCP} 
and isotropic antennas.
\citet{ramossomolinos2024} have determined the complex permittivity of lunar regolith at $f_\text{c} = \unit[2]{GHz}$ as $\epsilon_{\mathrm{g}} \approx 3.95 - 0.25\cj$.
If another surface communications system shall be considered, e.g. based on IEEE 802.11 or 3GPP standards, the cooperative navigation error model can easily be adapted for a different carrier frequency, bandwidth, waveform etc.

\begin{figure}[htb]
	\begin{minipage}{0.5\textwidth}
		\centering
		\includegraphics{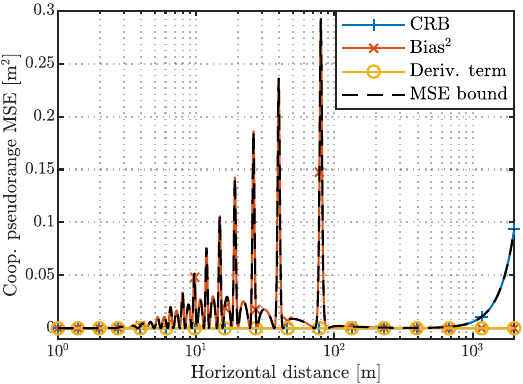}
		\caption{Contribution of the terms $\operatorname{CRB}(\hat{\rho})$, $\operatorname{B}^2(\hat{\rho})$ and $\operatorname{CRB}(\hat{\rho}) \left( 2 \nabla_\rho B(\hat{\rho}) + \nabla_\rho B^2(\hat{\rho}) \right)$ from \cref{eq:MSE} to the cooperative pseudorange \ac{MSE} $\operatorname{MSE}\{\hat{\rho}\}$ for $h_{\mathrm{Tx}} = \unit[6]{m}$ and $h_{\mathrm{Rx}} = \unit[1]{m}$}
		\label{fig:coopRangeMSE}
	\end{minipage}
	\hfill
	\begin{minipage}{0.5\textwidth}
		\centering
		\includegraphics{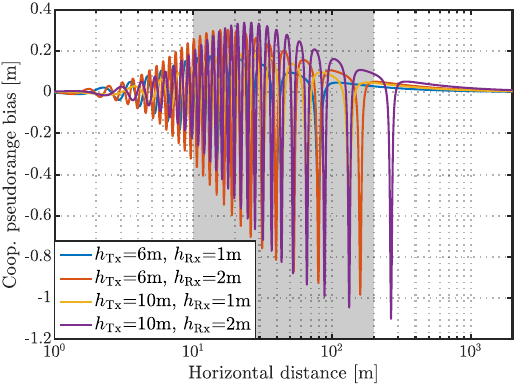}
		\caption{Cooperative pseudorange bias $B(d(d_{\mathrm{H}}))$ depending on horizontal distance $d_{\mathrm{H}}$ for different combinations of transmitter height $h_{\mathrm{Tx}}$ and receiver height $h_{\mathrm{Rx}}$}
		\label{fig:rangeBias_distance}
	\end{minipage}
\end{figure}

\Cref{fig:coopRangeMSE} shows the three additive terms from \cref{eq:MSE} individually and the resulting cooperative pseudorange estimation \ac{MSE} bound \cref{eq:MSE} for transmitter height $h_{\mathrm{Tx}} = \unit[6]{m}$ and receiver height $h_{\mathrm{Rx}} = \unit[1]{m}$.
We see that the bias is close to zero for small horizontal distances, highly variable for distances between \unit[5]{m} to \unit[100]{m} and goes to zero for large distances.
For distances above \unit[500]{m}, the estimation variance becomes the dominating error source.
The impact of the derivative term is negligible.
We thus set the cooperative pseudorange variance in \cref{eq:prCoopError} to $(\sigma_{\mathrm{ToF}}^k)^2 = \operatorname{CRB}(\hat{\rho})$,
which is calculated in closed-form \cref{eq:tofCRB} with the received signal power from the two-ray ground reflection model \cref{eq:Prx}.

Next, we focus on the cooperative pseudorange bias using the estimator \cref{eq:ml}.
\Cref{fig:rangeBias_distance} shows the cooperative pseudorange bias depending on the horizontal distance for
transmitter height \unit[6]{m} and \unit[10]{m}, representing the Argonaut lander with or without mast
and receiver height \unit[1]{m} and \unit[2]{m}, representing a
small lightweight rover and a large Curiosity-size rover, respectively.
By considering  minimum and maximum transmitter and receiver heights, we implicitly also cover different terrains.
From the plot, we see that for small and large distances, the pseudorange bias approaches zero.
For intermediate distances between \unit[3]{m} and \unit[300]{m}, the pseudorange bias varies quickly with horizontal distance.
With increasing transmitter/receiver height, the pseudorange bias amplitude increases.

In order to determine the exact pseudorange bias resulting from the ground reflection in the general three-dimensional case, we would need knowledge of
communications system parameters and antenna patterns,
ground electrical properties,
transmitter and receiver positions,
and ground reflection point position and slope, i.e. terrain.
Having accurate and global knowledge of all quantities that impact the pseudorange bias is not realistic in a practical scenario. Instead, we propose to model the average-case and worst-case cooperative pseudorange bias by a \ac{GMP-1}.

\subsection{Modeling Cooperative Pseudorange Bias as First Order Gauss-Markov Process}\label{ss:coopGMP}
In this paragraph, we start by defining the \ac{GMP-1} cooperative pseudorange bias model in discrete time domain, as required e.g. for use in a Kalman filter. In the next two paragraphs, we then show how to obtain the model parameters based on the cooperative pseudorange bias in distance domain determined in the last section.
Following \citet{kasdin1995}, we model the cooperative pseudorange bias for time step $k$ and time interval $T$ by a stationary discrete time \ac{GMP-1},
\begin{equation}\label{eq:coopGMP}
	b_{\mathrm{coop}}^k = e^{-\frac{T}{\tau_{\mathrm{coop}}}} \, b_{\mathrm{coop}}^{k-1} + u_{\rho_\mathrm{coop}}^k, \qquad
	u_{\rho_\mathrm{coop}}^k \sim \mathcal{N}\left(0,\, \sigma_{{\rho}_\mathrm{coop}}^2 (1 - e^{-2T/\tau_{\mathrm{coop}}})\right).
\end{equation}
According to this model, the cooperative pseudorange bias has
zero mean $\E{b_{\mathrm{coop}}^k} = \unit[0]{m}$,
variance $\var{b_{\mathrm{coop}}^k} = \sigma_{\mathrm{coop}}^2$ and
correlation time constant $\tau_{\mathrm{coop}}$.
For time $t$ and frequency $f$, the \ac{GMP-1} time domain \ac{ACF} and discrete time domain \ac{PSD} are given by
\begin{equation}
	R(t) = \sigma_{\mathrm{coop}}^2 e^{\frac{-\abs*{t}}{\tau_\mathrm{coop}}}, \qquad
	S(f) = \frac{\sigma_{{\rho}_\mathrm{coop}}^2 T \left(1 - e^{-\frac{2T}{\tau_{\mathrm{coop}}}}\right)}{1 + e^{-2 T/\tau_{\mathrm{coop}}} - 2e^{-T/\tau_{\mathrm{coop}}}\cos(2 \pi f T)}.\label{eq:coopACFPSD}
\end{equation}

\begin{figure}[htb]
	\begin{subfigure}{0.5\textwidth}
		\centering
		\includegraphics{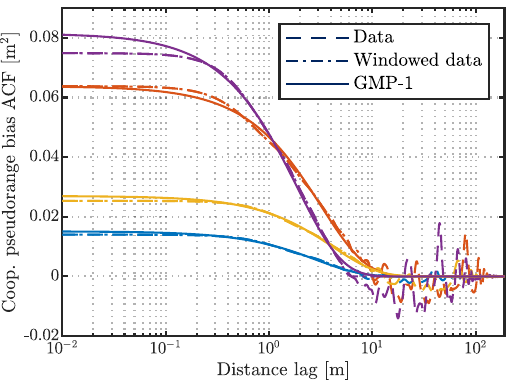}
		\caption{Cooperative pseudorange bias \acp{ACF}}
		\label{fig:ACFs_distance}
	\end{subfigure}
	\hfill
	\begin{subfigure}{0.5\textwidth}
		\centering
		\includegraphics{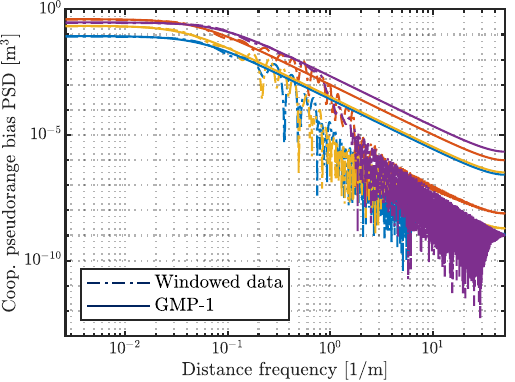}
		\caption{Cooperative pseudorange bias \acp{PSD}}
		\label{fig:PSDs_distance}
	\end{subfigure}
	\caption{Cooperative pseudorange bias \ac{ACF} and \ac{PSD} in horizontal distance domain based on data from \cref{fig:rangeBias_distance}, after applying a tapering window and the fitted \acp{GMP-1}}
	\label{fig:ACFsPSDs_distance}
\end{figure}

Having the defined the model, we determine the \ac{GMP-1} parameters from $N$ samples of empirical cooperative pseudorange bias data.
From \cref{fig:rangeBias_distance}, we have the
cooperative pseudorange bias $B(d(d_{\mathrm{H},n}))$ evaluated at discrete horizontal distances $d_{\mathrm{H},n}$ for $n = 1,...,N$,
where $d(d_{\mathrm{H}}) = \sqrt{(h_{\mathrm{Tx}} - h_{\mathrm{Rx}})^2 + d_{\mathrm{H}}^2}$.
We select the highlighted data part between \unit[10]{m} and \unit[200]{m} to calculate the sample \ac{ACF} in distance domain.
Then, we apply a tapering window suggested by \citet{langel2020} to the sample \ac{ACF} in order to limit spectral leakage.
We define a \ac{GMP-1} in discrete distance domain by replacing time and frequency in \cref{eq:coopACFPSD} by distance and distance frequency, respectively.
The respective discrete distance domain \ac{GMP-1} parameters are then estimated from the windowed sample \ac{ACF} $\hat{R}_n$ by
\begin{equation}
	\begin{bmatrix}
		\hat{\tau}_{\mathrm{d}} \\
		\hat{\sigma}_{\mathrm{coop}}^2
	\end{bmatrix} = 
	\min_{\tau_{\mathrm{d}},\sigma_{\mathrm{coop}}^2} \sum_{n=1}^N \abs*{\hat{R}_n - R(d_{\mathrm{H},n}, \tau_{\mathrm{d}}, \sigma_{\mathrm{coop}}^2)}^2,
\end{equation}
where $\hat{\tau}_{\mathrm{d}}$ is the distance correlation constant and $\hat{\sigma}^2$ is the variance.
From the four cooperative pseudorange bias curves in \cref{fig:rangeBias_distance}, we get four parameter sets $\{ \hat{\tau}_{\mathrm{d}}, \hat{\sigma}^2 \}$.
The four sample \acp{ACF}, windowed sample \acp{ACF}, and \ac{GMP-1} \acp{ACF} in horizontal distance domain are shown in \cref{fig:ACFs_distance}. Additionally, the \acp{PSD} obtained by taking the absolute \ac{FFT} of the windowed sample \acp{ACF} and the \ac{GMP-1} \acp{PSD} are shown in \cref{fig:PSDs_distance}.
As the \ac{GMP-1} \acp{ACF} match the windowed data well, we consider it a suitable model.
In \ac{PSD} domain, we see that the \ac{GMP-1} for large frequencies is greater than the windowed data, thus the model is conservative \citep{langel2020}.

\begin{figure}[htb]
	\begin{subfigure}{0.5\textwidth}
		\centering
		\includegraphics{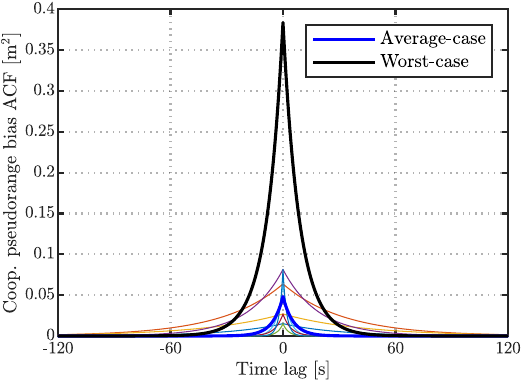}
		\caption{Cooperative pseudorange bias \acp{ACF}}
		\label{fig:ACFs_time}
	\end{subfigure}
	\hfill
	\begin{subfigure}{0.5\textwidth}
		\centering
		\includegraphics{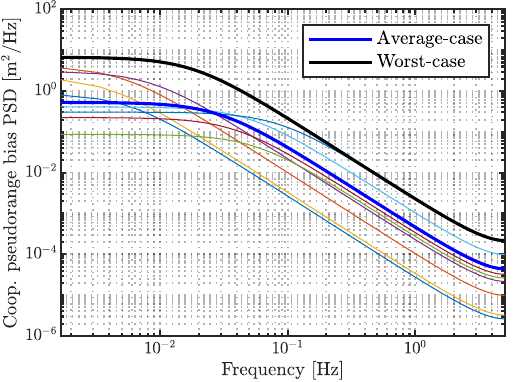}
		\caption{Cooperative pseudorange bias \acp{PSD}}
		\label{fig:PSDs_time}
	\end{subfigure}
	\caption{Cooperative pseudorange bias \acp{ACF} and \acp{PSD} of \acp{GMP-1} in time domain representing different transmitter and receiver heights and rover velocities (thin colored lines) as well as average and worst-case parameters}
	\label{fig:ACFsPSDs_time}
\end{figure}

To switch from horizontal distance to time domain, we additionally consider minimum and maximum rover radial velocities, $v_{\mathrm{min}}$ and $v_{\mathrm{max}}$. Non-radial motion of the rover is implicitly considered by $v_{\mathrm{min}}$. Exact tangential motion would result in $v_{\mathrm{min}} = \unitfrac[0]{m}{s}$, however this is unlikely in a real scenario, especially when considering terrain. We have chosen $v_{\mathrm{min}} = \unitfrac[0.1]{m}{s}$ and $v_{\mathrm{max}} = \unitfrac[1]{m}{s}$, which is considered very fast for planetary rovers \citep{debenedetti2024}.
Considering minimum and maximum rover radial velocity for all four cooperative pseudorange bias curves results in a total of eight parameter sets. Actually relevant are only
the minimum and maximum correlation distance constants $\hat{\tau}_{\mathrm{d,min}}$, $\hat{\tau}_{\mathrm{d,max}}$,
and the minimum and maximum variances $\sigma_{\mathrm{min}}^2$, $\sigma_{\mathrm{max}}^2$.
Following \citet{garciacrespillo2023}, where we use the simpler continuous-time equations as $T < \tau$, we obtain the worst-case \ac{GMP-1} parameters
\begin{equation}
	\hat{\tau}_{\mathrm{coop,worst}} = \sqrt{ (\hat{\tau}_{\mathrm{d,min}} / v_{\mathrm{max}}) \, (\hat{\tau}_{\mathrm{d,max}} / v_{\mathrm{min}}) }, \qquad
	\hat{\sigma}_{\mathrm{coop,worst}}^2 = \sqrt{\frac{\hat{\tau}_{\mathrm{d,max}} / v_{\mathrm{min}}}{\hat{\tau}_{\mathrm{d,min}} / v_{\mathrm{max}}}} \hat{\sigma}_{\mathrm{max}}^2,
\end{equation}
Similarly, we obtain average-case \ac{GMP-1} parameters
\begin{equation}
	\hat{\tau}_{\mathrm{coop,avg}}
	= \sqrt{ \left( \frac{(\hat{\tau}_{\mathrm{d,min}}  + \hat{\tau}_{\mathrm{d,max}})/2 }{(v_{\mathrm{min}} + v_{\mathrm{max}}/2 } \right)^2 }
	= \frac{\hat{\tau}_{\mathrm{d,min}} + \hat{\tau}_{\mathrm{d,max}}}{v_{\mathrm{min}} + v_{\mathrm{max}}}, \qquad
	\hat{\sigma}_{\mathrm{coop,avg}}^2 = \frac{\hat{\sigma}_{\mathrm{min}}^2 + \hat{\sigma}_{\mathrm{max}}^2}{2}.
\end{equation}
\Cref{fig:ACFs_time} shows the \ac{GMP-1} \acp{ACF} and
\cref{fig:PSDs_time} the \ac{GMP-1} \acp{PSD}
defined in \cref{eq:coopACFPSD}
for the eight individual parameter sets as well as the obtained average- and worst-case parameters.
In \ac{PSD} domain, it is apparent that the worst-case curve is an overbound to all other curves \citep{langel2020}.

The average- and worst-case \ac{GMP-1} parameters to model the cooperative pseudorange bias $b_{\mathrm{coop},l}^k$ in \cref{eq:prCoopError} by \cref{eq:coopGMP} are summarized in \cref{tab:CoopGMP}. This completes our cooperative navigation error model.

\begin{table}[htb]
	\caption{Cooperative pseudorange bias \ac{GMP-1} parameters}
	\label{tab:CoopGMP}
	\centering
	\begin{tabular}{lcc}
		\toprule
		& Average-case & Worst-case \\
		\midrule
		\ac{GMP-1} time constant $\tau_{\mathrm{coop}}$ & \unit[5.5]{s} & \unit[8.8]{s} \\
		\ac{GMP-1} standard deviation $\sigma_{\mathrm{coop}}$ & \unit[0.22]{m} & \unit[0.62]{m} \\
		\bottomrule
	\end{tabular}
\end{table}

\section{Satellite Navigation Error Model}\label{s:satNav}
\subsection{Pseudorange and Pseudorange Rate}
For the satellite navigation part, we consider pseudorange observations based on the signal delay, and pseudorange rate observations based on the signal Doppler between satellites $\mathds{S}$ and users $\mathds{U}$.
A pseudorange observation for a signal transmitted by satellite $j$ and received by user $i$ at epoch $k$ is defined as
\begin{equation}\label{eq:prSat}
	\rho_{i,j}^k = \norm{\bm{p}_j^k - \bm{p}_i^k} + c\delta_i^k - c\delta_j^k + \epsilon_{i,j}^{\rho,k},\: i \in \mathds{U},\: j \in \mathds{S},
\end{equation}
with user position $\bm{p}_i^k$, user clock offset $\delta_i^k$, satellite position $\bm{p}_i^k$ and satellite clock offset $\delta_i^k$.
The error $\epsilon_{i,j}^{\rho,k}$ is Gaussian distributed,
\begin{equation}\label{eq:prSatError}
	\epsilon_{i,j}^{\rho,k} \sim \mathcal{N}\left(b_{\mathrm{sat},j}^k,\, (\sigma_{\mathrm{DLL}}^k)^2\right),
\end{equation}
with mean $b_{\mathrm{sat},j}^k$ and variance $(\sigma_{\mathrm{DLL}}^k)^2$.
A pseudorange rate observation for a signal transmitted by satellite $j$ and received by user $i$ at epoch $k$ is defined as
\begin{equation}\label{eq:prrSat}
	\dot{\rho}_{i,j}^k = (\bm{v}_j^k - \bm{v}_i^k)^T \bm{u}_{i,j}^k + c\dot{\delta}_i^k - c\dot{\delta}_j^k + \epsilon_{i,j}^{\dot{\rho},k},\: i \in \mathds{U},\: j \in \mathds{S},
\end{equation}
with user velocity $\bm{v}_i^k$, user clock drift $\dot{\delta}_i^k$, satellite velocity $\bm{v}_j^k$, satellite clock drift $c\dot{\delta}_j^k$, unit vector from user $i$ to satellite $j$
\begin{equation}\label{eq:uijk}
	\bm{u}_{i,j}^k = \frac{\bm{p}_j^k - \bm{p}_i^k}{\norm{\bm{p}_j^k - \bm{p}_i^k}},
\end{equation}
and Gaussian error
\begin{equation}\label{eq:prrSatError}
	\epsilon_{i,j}^{\dot{\rho},k} \sim \mathcal{N}\left(\dot{b}_{\mathrm{sat},j}^k,\, (\sigma_{\mathrm{FLL}}^k)^2\right),
\end{equation}
with mean $\dot{b}_{\mathrm{sat},j}^k$ and variance $(\sigma_{\mathrm{FLL}}^k)^2$.
The noise variances $(\sigma_{\mathrm{DLL}}^k)^2$ and $(\sigma_{\mathrm{FLL}}^k)^2$ represent thermal noise of the \ac{DLL} and \ac{FLL}, respectively.
For carrier-to-noise ratio $\frac{C}{N_0}$,
carrier frequency $f_{\mathrm{c,sat}}$,
chip rate $f_{\mathrm{chip}}$,
\ac{DLL} loop bandwidth $B_{\mathrm{DLL}}$,
\ac{FLL} loop bandwidth $B_{\mathrm{FLL}}$,
coherent integration time $T_{\mathrm{i}}$,
and early-late spacing $d_{\mathrm{el}}$,
the noise variances are calculated following \citet{kaplan2006},
\begin{equation}
	(\sigma_{\mathrm{DLL}}^k)^2 = \frac{c^2}{f_{\mathrm{chip}}^2} \frac{B_{\mathrm{DLL}} d_{\mathrm{el}}}{2 \frac{C}{N_0}} \left(1+\frac{2}{T_{\mathrm{i}} \frac{C}{N_0} (2-d_{\mathrm{el}})}\right),
\end{equation}
\begin{equation}
	(\sigma_{\mathrm{FLL}}^k)^2 = \frac{c^2}{4 \pi^2 T_{\mathrm{i}}^2 f_{\mathrm{chip}}^2} \frac{4 B_{\mathrm{FLL}}}{\frac{C}{N_0}} \left(1+\frac{1}{T_{\mathrm{i}} \frac{C}{N_0}}\right).
\end{equation}
Neglecting multipath errors on the user side and due to the absence of a lunar atmosphere, the pseudorange bias $b_{\mathrm{sat},j}^k$ and pseudorange rate bias $\dot{b}_{\mathrm{sat},j}^k$ depend mainly on satellite orbit and clock errors, i.e. the \ac{SISE}. Clearly, the \ac{SISE} is satellite dependent and correlated over time \citep{stallo2023,gallon2022}. We thus model pseudorange bias and pseudorange rate bias as a discrete time stochastic process
\begin{equation}\label{eq:satGMP}
	\begin{bmatrix}
		b_{{\mathrm{sat},j}}^k \vspace{1mm}\\
		\dot{b}_{{\mathrm{sat},j}}^k \\
	\end{bmatrix} = 
	\bm{A}_{\mathrm{sat}}
	\begin{bmatrix}
		b_{\mathrm{sat},j}^{k-1} \vspace{1mm}\\
		\dot{b}_{\mathrm{sat},j}^{k-1} \\
	\end{bmatrix}
	+ \bm{u}_\mathrm{sat}^k
\end{equation}
with additive Gaussian noise $\bm{u}_\mathrm{sat}^k \sim \mathcal{N}(\bm{0}, \bm{U}_\mathrm{sat})$.
The process shall have zero mean $\E{b_{\mathrm{sat}}^k} = \unit[0]{m}$, $\E{\dot{b}_{\mathrm{sat}}^k} = \unitfrac[0]{m}{s}$
and variances $\var{b_{\mathrm{sat}}^k} = \sigma_{\rho_\mathrm{sat}}^2$, $\var{\dot{b}_{\mathrm{sat}}^k} = \sigma_{\dot{\rho}_\mathrm{sat}}^2$.
In the following, we discuss three choices of stochastic processes to model pseudorange bias and pseudorange rate bias.

\subsection{First Order Gauss-Markov Process (GMP-1)}\label{ss:GMP1}
\citet{gallon2022} have modeled GPS and Galileo orbit and clock errors by a stationary \ac{GMP-1}. They have used experimental data from the \ac{IGS} network to empirically determine the correlation time constant $\tau_{\mathrm{sat}}$ and the variance $\sigma_{{\rho}_\mathrm{sat}}^2$ characterizing the \ac{GMP-1}. However, they only consider pseudorange observations. We extend the model by an independent \ac{GMP-1} modeling pseudorange rate bias with the same time constant but different variance.
Following \citet{kasdin1995}, the \ac{GMP-1} model is defined by
\begin{equation}\label{eq:satGMP1}
	\bm{A}_{\mathrm{sat}} = \diag{\begin{bmatrix}
			e^{-\frac{T}{\tau_{\mathrm{sat}}}} & e^{-\frac{T}{\tau_{\mathrm{sat}}}}
	\end{bmatrix}},
\end{equation}
\begin{equation}\label{eq:U}
	\bm{U}_\mathrm{sat} = \diag{
		\begin{bmatrix}
			\sigma_{{\rho}_\mathrm{sat}}^2 (1 - e^{-2T/\tau_{\mathrm{sat}}}) & \sigma_{\dot{\rho}_\mathrm{sat}}^2 (1 - e^{-2T/\tau_{\mathrm{sat}}})
	\end{bmatrix}}.
\end{equation}
As can be seen from the diagonal nature of \cref{eq:satGMP1}, this model does not consider the correlation between pseudorange bias and pseudorange rate bias. Thus, the time derivative relationship between satellite position and velocity errors and satellite clock bias and drift, respectively, are ignored.

\subsection{Integrated First Order Gauss-Markov Process (IGMP-1)}\label{ss:IGMP1}
The first idea to consider a time derivative relationship might be to model the pseudorange bias as \ac{GMP-1} \citep{gallon2022} and the pseudorange rate bias as its derivative. However, the derivative of a continuous time \ac{GMP-1} would have infinite variance \citep{brown2012}. Although we consider discrete time processes in this work, they stem from sampling their continuous time counterparts. Thus, the model should not be violated in continuous time.

Instead, we can model the pseudorange rate bias as \ac{GMP-1} and the pseudorange bias as the respective time integral. The resulting \ac{IGMP-1} is defined as
\begin{equation}\label{eq:satIGMP1}
	\bm{A}_{\mathrm{sat}} = \begin{bmatrix}
		1 & \tau_{\mathrm{sat}}\left(1-e^{-\frac{T}{\tau_{\mathrm{sat}}}}\right) \\
		0 & e^{-\frac{T}{\tau_{\mathrm{sat}}}}
	\end{bmatrix},
\end{equation}
where $\alpha_{\mathrm{sat}}$ is defined in \cref{eq:satGMP1} and for $T/\tau_{\mathrm{sat}} \ll 1$ the covariance matrix is given by \citet{bar-shalom2004},
\begin{equation}
	\bm{U}_\mathrm{sat} \approx \frac{2 \sigma_{\dot{\rho}_\mathrm{sat}}^2}{\tau_{\mathrm{sat}}}
	\begin{bmatrix}
		T^3/3 & T^2/2 \\
		T^2/2 & T
	\end{bmatrix}.
\end{equation}
The downside of this model is that it is not stationary and the pseudorange bias $b_{{\mathrm{sat}}}^k$ grows without bounds. Still, we believe this is a suitable model if ephemeris updates, where satellite orbit and clock errors are reset to a low value, shall be considered explicitly. If modeling ephemeris update intervals is out of scope, other models might be more suitable.

\subsection{Second Order Gauss-Markov Process (GMP-2)}\label{ss:GMP2}
\citet{leonard2013} have proposed a \ac{GMP-2} to model satellite orbit dynamics.
From \citet{bryson1975} we find the transition matrix for a stationary \ac{GMP-2},
\begin{equation}\label{eq:satGMP2}
	\bm{A}_{\mathrm{sat}} = 
	e^{-\zeta \omega T}
	\begin{bmatrix}
		\cos(\beta T) + \frac{\zeta\omega}{\beta}\sin(\beta T) & \frac{1}{\beta} \sin(\beta T) \\
		-\frac{\omega^2}{\beta}\sin{\beta T} & \cos(\beta T)-\frac{\zeta\omega}{\beta}\sin(\beta T)
	\end{bmatrix},
\end{equation}
with natural frequency $\omega = 1/\tau_{\mathrm{sat}}$,
damping coefficient $\zeta$ and $\beta = \omega\sqrt{1-\zeta^2}$. The noise covariance matrix $\bm{U}_\mathrm{sat}$ can be computed numerically by the method from \citet{vanloan1978}.
We have determined $\zeta \approx 0.7$ numerically by matching the \acp{ACF} of \ac{GMP-2} to two \acp{GMP-1} in a mean square sense.

In the next section, we determine suitable model parameters for lunar \ac{PNT} and compare the three models by their \acp{ACF}. In \cref{s:case} we compare models in terms of the resulting \ac{BCRB} for position estimation.

\subsection{Parameters and Comparison of Satellite Navigation Error Models}\label{ss:satParams}
As lunar navigation satellites are not yet in place, determining parameters from measurement data is not possible.
We thus use the correlation time constant $\tau_{\mathrm{sat}} = \unit[5]{h}$ determined by \citet{gallon2022} for GPS satellites with Rubidium clocks. It is clear, that the correlation time constant of lunar satellites might be different. However, in our analysis, we saw the importance of having the correct order of magnitude of the correlation time constant. Variations within one order of magnitude only had a small impact on the resulting position accuracy.
Once data from high fidelity simulations or measurement data is available, the model parameters can be updated.

As worst-case pseudorange bias standard deviation, we use $\sigma_{\rho_{\mathrm{sat}},\mathrm{worst}} = \unit[10]{m}$ following the \unit[20]{m} $95\%$ \ac{SISE} target by ESA \citep{ventura-traveset2024}. The achievability of the target is confirmed by \citet{stallo2023} for maximum \unit[6]{h} ephemeris age of data. 
The related worst-case pseudorange rate bias standard deviation $\sigma_{\dot{\rho}_{\mathrm{sat}},\mathrm{worst}} = \sigma_{\rho_{\mathrm{sat}},\mathrm{worst}} / \tau_{\mathrm{sat}} = \unitfrac[0.56]{mm}{s}$ is determined from the continuous time \ac{GMP-2} stationary covariance matrix \citep{bryson1975}.
As average-case pseudorange bias standard deviation, we assume $\sigma_{\rho_{\mathrm{sat}},\mathrm{avg}} = \unit[5]{m}$ based on the analysis by \citet{stallo2023} for \unit[2]{h} ephemeris age of data. %
The average-case pseudorange rate bias standard deviation is again defined as
$\sigma_{\dot{\rho}_{\mathrm{sat}},\mathrm{avg}} = \sigma_{\rho_{\mathrm{sat}},\mathrm{avg}} / \tau_{\mathrm{sat}} = \unitfrac[0.28]{mm}{s}$.

\begin{figure}[htb]
	\begin{subfigure}{0.5\textwidth}
		\centering
		\includegraphics{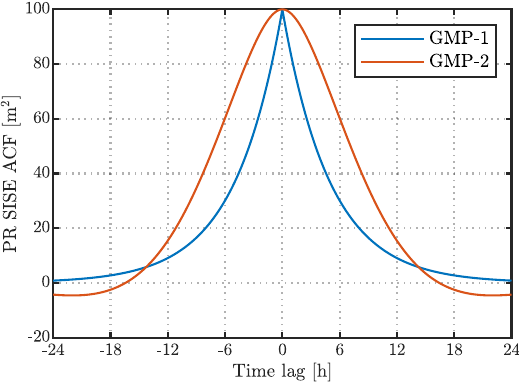}
		\caption{Pseudorange \ac{SISE} \ac{ACF}}
		\label{fig:PR_SISE_ACF}
	\end{subfigure}
	\hfill
	\begin{subfigure}{0.5\textwidth}
		\centering
		\includegraphics{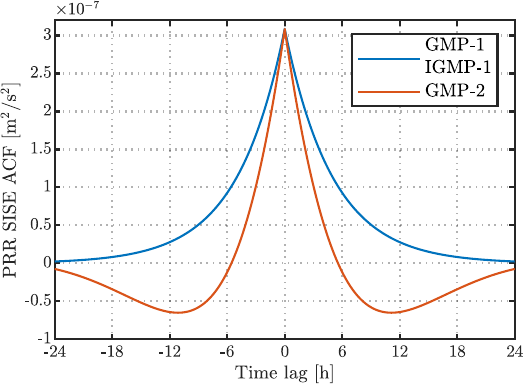}
		\caption{Pseudorange rate \ac{SISE} \ac{ACF}}
		\label{fig:PRR_SISE_ACF}
	\end{subfigure}
	\caption{Satellite navigation pseudorange and pseudorange rate bias \acp{ACF} for different Gauss-Markov processes}
	\label{fig:SISE_ACFs}
\end{figure}

\Cref{fig:PR_SISE_ACF} shows the pseudorange bias \ac{ACF} and \cref{fig:PRR_SISE_ACF} the pseudorange rate bias \acp{ACF} of the different models for the defined worst-case parameters.
As apparent from \cref{eq:satGMP1,eq:satIGMP1}, \ac{GMP-1} and \ac{IGMP-1} share the same pseudorange rate bias \acp{ACF}.
For \ac{IGMP-1}, no stationary pseudorange bias \acp{ACF} exists.
\ac{GMP-2} and \ac{GMP-1} are different stochastic processes with distinct properties. Their \acp{ACF} cannot be matched exactly.
Comparing the mainlobes of the pseudorange bias \acp{ACF}, \ac{GMP-2} has a wider mainlobe, indicating stronger short-term correlation.

\section{Hybrid Navigation System Model}\label{s:hybrid}
\subsection{Augmented State Space}
The state of an individual user $i$ is defined as
\begin{equation}\label{eq:userState}
	\bm{\tilde{x}}_{i}^k =
	\begin{bmatrix}
		(\bm{p}_i^k)^T & (\bm{v}_i^k)^T & c\delta_i^k & c\dot{\delta}_i^k
	\end{bmatrix}^T,
\end{equation}
with the three-dimensional user position $\bm{p}_i^k$ and velocity $\bm{v}_i^k$. The user clock offset $\delta_i^k$ and clock drift $\dot{\delta}_i^k$ are multiplied by the speed of light $c$.
The state of all users 
\begin{equation}\label{eq:x}
	\bm{\tilde{x}}^k = 
	\begin{bmatrix}
		(\bm{\tilde{x}}_1^k)^T & ... & (\bm{\tilde{x}}_i^k)^T & ... & (\bm{\tilde{x}}_{|\mathds{U}|}^k)^T
	\end{bmatrix}^T
\end{equation}
is obtained by stacking the states of $|\mathds{U}|$ users.
We further define the set of cooperative links $\mathds{L}_{\mathrm{coop}} = \{ ..., l(i,j), ... \}\, \forall i,j \in \mathds{U}$.
Based on \cref{s:coopNav,s:satNav}, we model satellite pseudorange bias $b_{\mathrm{sat},j}^k$ and satellite pseudorange rate bias $\dot{b}_{\mathrm{sat},j}^k$ for all satellites $j \in \mathds{S}$ as well as cooperative pseudorange bias $b_{\mathrm{coop},l}^k$ for all cooperative links $l \in \mathds{L}_{\mathrm{coop}}$.
Including
\begin{equation}
	\bm{b}^k = 
	\begin{bmatrix}
		\hdots & b_{\mathrm{sat},j}^k & \dot{b}_{\mathrm{sat},j}^k & \hdots & b_{\mathrm{coop},l}^k \hdots
	\end{bmatrix}^T, j \in \mathds{S},\, l \in \mathds{L}_{\mathrm{coop}}
\end{equation}
into the state space, we obtain the augmented state space model
\begin{equation}\label{eq:augTransition}
	\underbrace{\begin{bmatrix}
			\bm{\tilde{x}}^k \\
			\bm{b}^k
	\end{bmatrix}}_{\bm{x}^k} = 
	\underbrace{\begin{bmatrix}
			\bm{\tilde{F}}^k & \mymathbb{0} \\
			\mymathbb{0} & \bm{A}^k
	\end{bmatrix}}_{\bm{F}^k}
	\underbrace{\begin{bmatrix}
			\bm{\tilde{x}}^{k-1} \\
			\bm{b}^{k-1}
	\end{bmatrix}}_{\bm{x}^{k-1}} +
	\underbrace{\begin{bmatrix}
			\bm{\tilde{D}}^{k} \bm{\tilde{o}}^{k} \\
			\bm{0}
	\end{bmatrix}}_{\bm{d}^k} +
	\underbrace{\begin{bmatrix}
			\bm{\tilde{w}}^k \\
			\bm{u}^k
	\end{bmatrix}}_{\bm{w}^k},
\end{equation}
with transition matrices $\bm{\tilde{F}}^k$, $\bm{A}^k$ and control vector $\bm{\tilde{o}}^{k}$, mapped to the state space by matrix $\bm{\tilde{D}}^{k}$.
We define $\mymathbb{0}$ as an all zero matrix and $\bm{0}$ as an all zero column vector of appropriate dimension.
Furthermore, $\bm{\tilde{w}}^k$ and $\bm{u}^k$ are process noise vectors with the augmented process noise covariance matrix
\begin{equation}
	\bm{Q} = \E{
		\begin{bmatrix}
			\bm{\tilde{w}}^k \\
			\bm{u}^k
		\end{bmatrix}
		\begin{bmatrix}
			(\bm{\tilde{w}}^k)^T & (\bm{u}^k)^T
	\end{bmatrix}} =
	\begin{bmatrix}
		\bm{\tilde{Q}} & \mymathbb{0} \\
		\mymathbb{0} & \bm{U}
	\end{bmatrix}.
\end{equation}

\subsection{Process Models}
The transition matrix of the user state $\bm{\tilde{F}}^{k} = \diag{\hdots,\, \bm{\tilde{F}}_i^{k},\, \hdots}$
is block-diagonal, with the block corresponding to user $i$ given by
\begin{equation}\label{eq:Fik}
	\bm{\tilde{F}}_i^{k} = \begin{bmatrix}
		\mathbb{I}_3 & T\mathbb{I}_3 & \bm{0} & \bm{0} \\
		\mymathbb{0}_3 & \mathbb{I}_3 & \bm{0} & \bm{0} \\
		\bm{0}^T & \bm{0}^T & 1 & T \\
		\bm{0}^T & \bm{0}^T & 0 & 1
	\end{bmatrix},
\end{equation}
where $T$ is the time interval between epochs, $\mathbb{I}_n$ is an identity matrix of dimension $n$.
We define $\diag{.}$ as an operator that creates a square matrix from the elements or sub-matrices on the diagonal.
A continuous white noise acceleration model \citep{bar-shalom2004} is assumed for the position and velocity states and a two-state clock model for the clock states.
The control vector $\bm{\tilde{o}}^k$ and matrix $\bm{\tilde{D}}^k$ for each user $i$ are composed of $\bm{\tilde{o}}_i^k$ and $\bm{\tilde{D}}_i^k$ analogous to the transition matrix $\bm{\tilde{F}}^{k}$.
For static users, $\bm{\tilde{o}}_i^k$ is a zero vector.
Assuming moving users are rovers with velocity control, we have
$\bm{D}_i^k = \begin{bmatrix}
	\mymathbb{0}_{3\times3} &
	\mathbb{I}_3 &
	\mymathbb{0}_{2\times3} 
\end{bmatrix}^T$.
The user process noise is Gaussian distributed,
$\bm{\tilde{w}}^k \sim \mathcal{N}(\bm{0},\bm{\tilde{Q}})$,
with a block-diagonal process noise covariance matrix
$\bm{\tilde{Q}} = \diag{\hdots,\, \bm{\tilde{Q}}_i,\, \hdots}$
defined per user $i$,
\newcommand{\sigmaiv}{\sigma_{\mathrm{v},i}}
\newcommand{\sigmaic}[1]{\sigma_{\mathrm{c}#1,i}}
\begin{equation}\label{eq:Qik}
	\bm{\tilde{Q}}_i = \begin{bmatrix}
		\frac{T^3 \sigmaiv^2}{3} \mathbb{I}_3 & \frac{T^2 \sigmaiv^2}{2} \mathbb{I}_3 & \bm{0} & \bm{0} \\
		\frac{T^2 \sigmaiv^2}{2} \mathbb{I}_3 & T \sigmaiv^2 \mathbb{I}_3 & \bm{0} & \bm{0} \\
		\bm{0}^T & \bm{0}^T & \sigmaic{1}^2 T + \sigmaic{2}^2 \frac{T^3}{3} & \sigmaic{2}^2 \frac{T^2}{2} \\
		\bm{0}^T & \bm{0}^T & \sigmaic{2}^2 \frac{T^2}{2} & \sigmaic{2}^2 T
	\end{bmatrix},
\end{equation}
with velocity noise coefficient $\sigmaiv$ and clock noise coefficients $\sigmaic{1}$, $\sigmaic{2}$. For static users, the $\bm{v}_i^k$ is omitted from the state vector and $\sigmaiv = \unit[0]{m/s^{1.5}}$.

The state transition of the augmented part is given by
\begin{equation}\label{eq:Ak}
	\bm{A}^k = \diag{\begin{bmatrix}
			\hdots & \bm{A}_{\mathrm{sat},j} & \hdots & \alpha_{\mathrm{coop},l} & \hdots
	\end{bmatrix}}, j \in \mathds{S},\, l \in \mathds{L}_{\mathrm{coop}},
\end{equation}
where the transition of the satellite bias state $\bm{A}_{\mathrm{sat},j}$ is defined by \cref{eq:satGMP1} for \ac{GMP-1}, \cref{eq:satIGMP1} for \ac{IGMP-1} and \cref{eq:satGMP2} for \ac{GMP-2}, respectively. For the cooperative counterpart, $\alpha_{\mathrm{coop},l}$ is defined by \cref{eq:coopGMP}.
The process noise of the augmented part is also Gaussian, $\bm{u}^k \sim \mathcal{N}(\bm{0},\bm{U})$, 
with the process noise covariance matrix
\begin{equation}
	\bm{U} = \diag{
		\begin{bmatrix}
			\hdots &
			\bm{U}_{\mathrm{sat},j} &
			\hdots & \sigma_{{\rho}_\mathrm{coop}}^2 (1 - e^{-2T/\tau_{\mathrm{coop}}}) & \hdots
	\end{bmatrix}}, j \in \mathds{S},\, l \in \mathds{L}_{\mathrm{coop}}.
\end{equation}
The appropriate process noise covariance matrix $\bm{U}_{\mathrm{sat},j}$ for the satellite part is chosen from \ac{GMP-1}, \ac{IGMP-1} or \ac{GMP-2} and the process noise variance for the cooperative part is defined according to \cref{eq:coopGMP}.

\subsection{Observation Models}
The observation model is
\begin{equation}\label{eq:z}
	\bm{z}^k = \bm{h}(\bm{x}^k) + \bm{r}^k, \qquad
	\bm{h}(\bm{x}^k) =
	\begin{bmatrix}
		\hdots &
		\rho_{i,j}^k &
		\dot{\rho}_{i,j}^k &
		\hdots
		\rho_{i,m}^k &
		\hdots
	\end{bmatrix}^T,\, i,m \in \mathds{U},\, j \in \mathds{S},
\end{equation}
where
satellite pseudorange observations $\rho_{i,j}^k$ for user $i$ and satellite $j$ are defined by \cref{eq:prSat},
satellite pseudorange rate observations $\dot{\rho}_{i,j}^k$ by \cref{eq:prrSat},
and cooperative pseudorange observations $\rho_{i,m}^k$ for receiving user $i$ and transmitting user $m$ by \cref{eq:prCoop}.
The observation noise is Gaussian distributed, $\bm{r}^k \sim \mathcal{N}(\bm{0}, \bm{R}^k)$,
with observation covariance matrix
\begin{equation}\label{eq:Rk}
	\bm{R}^k = \diag{...,\, \sigma_{\rho_{i,j}^k}^2,\, \sigma_{\dot{\rho}_{i,j}^k}^2,\, ...,\, \sigma_{\dot{\rho}_{i,m}^k}^2,\, ...},\: i,m \in \mathds{U},\, j \in \mathds{S}.
\end{equation}

For the calculation of the Bayesian Cramér-Rao bound in \cref{s:bcrb} and the hybrid navigation algorithms in \cref{s:algorithms}, we also need the observation Jacobian
\begin{equation}
	\bm{H}^k = 
	\frac{\partial\bm{h}(\bm{x}^k)}{\partial\bm{x}^k}
	=
	\begin{bmatrix}
		\bm{\tilde{H}}^k & \bm{E}^k
	\end{bmatrix},
\end{equation}
representing the linearization of the observation model about the state.
The Jacobian of the full user state $\bm{\tilde{H}}^k$ is further partitioned into
\begin{equation}\label{eq:Hk}
	\bm{\tilde{H}}^k = \begin{bmatrix}
		& \vdots & & \vdots \\		
		\hdots & \bm{\tilde{H}}_{l(i,j),i}^k & \hdots & \bm{\tilde{H}}_{l(i,j),j}^k \hdots \\
		& \vdots & & \vdots \\
	\end{bmatrix},
\end{equation}
where the block-columns refer to the receiving user $i \in \mathds{U}$ and transmitting satellite or user $j \in \mathds{S} \cup \mathds{U}$ and the block-rows refer to the index of the particular link $l(i,j)$, respectively.
The content of the blocks $\bm{\tilde{H}}_{l(i,j),i}^k$ depends on whether the signal comes from a satellite or a neighboring user and on the cooperation mode.
In this paper, we assume physical layer cooperation, where the cooperative navigation subsystem, see \cref{ss:coopPRError}, and the satellite navigation subsystems are driven by the same oscillator. This allows to use pseudorange observations among users \cref{eq:prCoop} together with pseudorange observations from satellites \cref{eq:prSat} in a unified system model.
Furthermore, \citet{pohlmann2024} have shown that physical layer cooperation provides superior performance compared to localization layer cooperation, where cooperative ranging is provided by an independent subsystem and users only observe noisy distances to their neighbors.
For signals transmitted by satellites, we consider pseudorange and pseudorange rate observations, for signals transmitted by users we consider only pseudorange observations. The respective Jacobian blocks are thus
\begin{equation}
	\bm{\tilde{H}}_{l(i,j),i}^k = -\bm{\tilde{H}}_{l(i,j),j}^k =
	\begin{cases}
		\begin{bmatrix}
			-(\bm{u}_{i,j}^k)^T & \bm{0}^T & 1 & 0 \vspace{0.5ex}\\
			-(\bm{v}_{i,j}^k)^T & -(\bm{u}_{i,j}^k)^T & 0 & 1
		\end{bmatrix}, & j \in \mathds{S}\vspace{1ex}\\
		\begin{bmatrix}
			-(\bm{u}_{i,j}^k)^T & \bm{0}^T & 1 & 0 \vspace{0.5ex}
		\end{bmatrix}, & j \in \mathds{U}.
	\end{cases}
\end{equation}
The unit vector $\bm{u}_{i,j}^k$ is given by \cref{eq:uijk} and%
\begin{equation}
	\bm{v}_{i,j}^k = \frac{(\mathbb{I}_3 - \bm{u}_{i,j}^k (\bm{u}_{i,j}^k)^T) (\bm{v}_j^k - \bm{v}_i^k)}{\norm{\bm{p}_j^k - \bm{p}_i^k}}.
\end{equation}
Defining $o(i,j)$ as a function returning the index within observation vector \cref{eq:z} corresponding to a signal transmitted by $j$ and received by $i$, the Jacobian of the augmented part, $\bm{E}^k$, consists of all zeros and ones at
\begin{equation}
	\begin{split}
		&[\bm{E}^k]_{o(i, j),2j-1} = [\bm{E}^k]_{o(i, j),2j} = 1\: \forall\, i \in \mathds{U},\, j \in \mathds{S}, \\
		&[\bm{E}^k]_{o(i, m),2\abs{\mathds{S}}+l} = 1,\, \forall\, i,m \in \mathds{U},\, l \in \mathds{L}_{\mathrm{coop}}.
	\end{split}
\end{equation}

\section{Bayesian Cramér-Rao Bound}\label{s:bcrb}
As a fundamental limit for hybrid lunar \ac{PNT} performance, we calculate the recursive Bayesian Cramér-Rao bound for tracking following \citet{tichavsky1998,vantrees2007}. We start with the prior \ac{BIM} $\bm{J}^0 = ( \bm{\Sigma}^0 )^{-1}$, which is defined with
\begin{equation}\label{eq:Sigma0}
	\bm{\Sigma}^0 =
	\begin{bmatrix}
		\bm{\tilde{\Sigma}}^0 & \bm{0} \\
		\bm{0} & \bm{U}
	\end{bmatrix}, \qquad
	\bm{\tilde{\Sigma}}^0 = \diag{\begin{bmatrix}
			\hdots & \sigma_{\bm{p}_i^0}^2 \mathbb{I}_3 & \sigma_{\bm{v}_i^0}^2 \mathbb{I}_3 & \sigma_{c\delta_i^0}^2 & \sigma_{c\dot{\delta}_i^0}^2 & \hdots
	\end{bmatrix}},\, i \in \mathds{U},
\end{equation}
and $\bm{U}$ from \cref{eq:U}.
As the process model \cref{eq:augTransition} is linear and the observation model \cref{eq:z} is nonlinear, the \ac{BIM} for consecutive epochs is calculated recursively by
\begin{equation}\label{eq:BIM}
	\bm{J}^k = \left(\bm{Q}^{k-1} + \bm{F}^{k-1} (\bm{J}^{k-1})^{-1} (\bm{F}^{k-1})^T\right)^{-1}
	+ \EE{\bm{x}^k}{(\bm{H}^k)^T (\bm{R}^k)^{-1} \bm{H}^k}.
\end{equation}
While \cref{eq:BIM} is exact, we can neglect the expectation operator and evaluate the equation only at the true state \citep{vantrees2007} due to the small process noise for the target application.
The \ac{BCRB} for epoch $k$ is then obtained by inverting the \ac{BIM},
\begin{equation}
	\MSE{\bm{x}^k} \geq \operatorname{BCRB}\left(\bm{x}^k\right) = \left(\bm{J}^k\right)^{-1}.
\end{equation}
Since the estimated user positions are influenced by all other states, the main quantity of interest is the mean position error bound of all users in the network,
\begin{equation}\label{eq:peb}
	\sqrt{\frac{1}{|\mathds{U}|} \sum_{i \in \mathds{U}} \trace{\operatorname{MSE}\{\bm{\hat{p}}_i^k\}}} \geq	
	\sqrt{\frac{1}{|\mathds{U}|} \sum_{i \in \mathds{U}} \trace{\mathrm{BCRB}(\bm{p}_i^k)}}.
\end{equation}

\section{Hybrid Navigation Algorithms}\label{s:algorithms}
We consider three hybrid navigation algorithms. All of them are centralized, meaning all observations of the users are available at a centralized entity, e.g. the lander. How hybrid lunar \ac{PNT} can be achieved in a distributed fashion has been investigated by \citet{pohlmann2025a} and is not within the scope of this paper. The algorithms operate on the augmented state space \cref{eq:augTransition}.

\subsection{Augmented \Acf{EKF}}
The first hybrid navigation algorithm is based on the \ac{EKF}, which is commonly used in navigation. It follows a Gaussian assumption of process and observation noise. Thus, the uncertainty of the estimated state $\bm{x}^k$ for epoch $k$ is captured by the covariance matrix $\bm{\Sigma}^k$.
The \ac{EKF} consists of the two steps prediction based on the process model and update based on the observation model.
Since the process model \cref{eq:augTransition} is linear, we use the prediction equations of the standard Kalman filter,
\begin{align}\label{eq:EKFprediction}
	\bm{x}_{\mathrm{pred}}^k &= \bm{F}^k\bm{x}^{k-1} + \bm{d}^{k}, \\
	\bm{\Sigma}_{\mathrm{pred}}^k &= \bm{F}^k\bm{\Sigma}^{k-1}(\bm{F}^k)^T + \bm{Q}^{k}.
\end{align}
For the nonlinear observation model \cref{eq:z}, we use the linearized \ac{EKF} update,
\begin{align}\label{eq:EKFupdate}
	\bm{x}^k &= \bm{x}_{\mathrm{pred}}^k + \bm{K}^k \left(\bm{z}^k - \bm{h}(\bm{x}_{\mathrm{pred}}^k)\right), \\
	\bm{\Sigma}^k &= \left(\mathbb{I} - \bm{K}^k \bm{H}^k\right) \bm{\Sigma}_{\mathrm{pred}}^k \left(\mathbb{I} - \bm{K}^k \bm{H}^k\right)^T + \bm{K}^k \bm{R}^k (\bm{K}^k)^T, \\
	\bm{K}^k &= \bm{\Sigma}_{\mathrm{pred}}^k(\bm{H}^k)^T \left(\bm{H}^k\bm{\Sigma}_{\mathrm{pred}}^k(\bm{H}^k)^T + \bm{R}^k\right)^{-1},
\end{align}
where the observation Jacobian is evaluated at the predicted state, $\bm{H}^k = \frac{\partial\bm{h}(\bm{x}^k)}{\partial\bm{x}^k} \big|_{\bm{x}^k = \bm{x}_{\mathrm{pred}}^k}$.
We use the covariance update equation by \citet{bucy1968} to ensure the covariance matrix $\bm{\Sigma}^k$ is symmetric positive definite. For the first step, the prior covariance matrix is given by \cref{eq:Sigma0}.

\subsection{Augmented Iterated Extended Kalman Filter (IEKF)}
The update step of the \ac{EKF} is based on a first order Taylor expansion about the predicted state. When the observation model is highly nonlinear, this can introduce errors, leading to suboptimal performance. One approach towards improvement is the \ac{IEKF} \citep{simon2006}. It is based upon the idea that after the update step, we have a better estimate of the state available, which is a better point to perform the Taylor expansion. Thus, when executing the \ac{IEKF}, the observation model is re-linearized several times until convergence.
The prediction step is identical to the \ac{EKF} \cref{eq:EKFprediction}.
The initial update step for $n=0$ to obtain $\bm{x}_0^k$ and $\bm{\Sigma}_0^k$ is also identical to the \ac{EKF} \cref{eq:EKFupdate}. Then, for subsequent update steps $n=1,2,...$ we iterate
\begin{align}
	\bm{x}_n^k &= \bm{x}_{\mathrm{pred}}^k + \bm{K}_n^k \left(\bm{z}^k - \bm{h}(\bm{x}_{\mathrm{pred}}^k) - \bm{H}_n^k\left(\bm{x}_{\mathrm{pred}}^k - \bm{x}_{n-1}^k\right) \right),  \\
	\bm{\Sigma}_n^k &= \left(\mathbb{I} - \bm{K}_n^k \bm{H}_n^k\right) \bm{\Sigma}_{\mathrm{pred}}^k \left(\mathbb{I} - \bm{K}_n^k \bm{H}_n^k\right)^T + \bm{K}_n^k \bm{R}^k (\bm{K}_n^k)^T, \\
	\bm{K}_n^k &= \bm{\Sigma}_{\mathrm{pred}}^k(\bm{H}_n^k)^T \left(\bm{H}_n^k\bm{\Sigma}_{\mathrm{pred}}^k(\bm{H}_n^k)^T + \bm{R}^k\right)^{-1},
\end{align}
until convergence. At every iteration step, the observation Jacobian $\bm{H}_n^k$ is calculated based on the state estimate from the previous iteration step $\bm{H}_n^k = \frac{\partial \bm{h}(\bm{x}^k)}{\partial \bm{x}^k} \big|_{\bm{x}^k = \bm{x}_{n-1}^k}$. The price to pay for the expected improved performance for highly nonlinear observation models is the increased computational complexity compared to the \ac{EKF}. For small to medium state dimension this is typically not an issue, however it could become prohibitive for large state dimension.

\subsection{Augmented Second Order Extended Kalman Filter (EKF-2)}
Instead of a first order Taylor expansion about the predicted state as for the \ac{EKF}, we can also perform a second order Taylor expansion, leading to the \ac{EKF-2}. We follow the derivation by \citet{roth2011}, which considers the second order Taylor expansion for both, state and covariance update, in contrast to other simplified versions.
The prediction step is again identical to the \ac{EKF} \cref{eq:EKFprediction}.
For the update step, in addition to the observation Jacobian $\bm{H}^k = \frac{\partial\bm{h}(\bm{x}^k)}{\partial\bm{x}^k} \big|_{\bm{x}^k = \bm{x}_{\mathrm{pred}}^k}$, we also need the observation Hessian $\bm{N}_o^k$ of each observation vector element $o$,
\begin{equation}\label{eq:N}
	\left[\bm{N}_o^k\right]_{p,q} = \frac{\partial^2 \left[\bm{h}\left(\bm{x}^k\right)\right]_o}{\partial [\bm{x}^k]_p \partial [\bm{x}^k]_q} \Bigg|_{\bm{x}^k = \bm{x}_{\mathrm{pred}}^k}.
\end{equation}
With the predicted observations
\begin{equation}\label{eq:sPred}
	\left[\bm{z}_{\mathrm{pred}}^k\right]_o = \left[\bm{h}\left(\bm{x}_{\mathrm{pred}}^k\right)\right]_o + \frac{1}{2} \trace{\bm{N}_o^k \bm{\Sigma}_{\mathrm{pred}}^k},
\end{equation}
and the second order covariance term
\begin{equation}\label{eq:S}
	\left[\bm{S}^k\right]_{l,m} = \frac{1}{2} \trace{\bm{N}_l^k \bm{\Sigma}_{\mathrm{pred}}^k \bm{N}_o^k \bm{\Sigma}_{\mathrm{pred}}^k},
\end{equation}
the update of the \ac{EKF-2} is performed as
\begin{align}
	\bm{x}^k &= \bm{x}_{\mathrm{pred}}^k + \bm{K}^k \left(\bm{z}^k - \bm{z}_{\mathrm{pred}}^k\right), \\
	\bm{\Sigma}^k &= \left(\mathbb{I} - \bm{K}^k \bm{H}^k\right) \bm{\Sigma}_{\mathrm{pred}}^k \left(\mathbb{I} - \bm{K}^k \bm{H}^k\right)^T + \bm{K}^k \left( \bm{R}^k + \bm{S}^k \right) (\bm{K}^k)^T, \\
	\bm{K}^k &= \bm{\Sigma}_{\mathrm{pred}}^k (\bm{H}^k)^T \left(\bm{H}^k\bm{\Sigma}_{\mathrm{pred}}^k(\bm{H}^k)^T + \bm{R}^k + \bm{S}^k\right)^{-1}.
\end{align}
As can be seen from \cref{eq:N,eq:sPred,eq:S}, the complexity of \ac{EKF-2} is significantly higher compared to the \ac{EKF}. Nevertheless, it can be applied for small to medium state and observation vector dimension.

\section{Lunar Navigation Case Studies}\label{s:case}
\subsection{Scenario}\label{ss:scenario}
\begin{figure}[htb]
	\begin{minipage}{0.45\textwidth}
		\centering
		\includegraphics{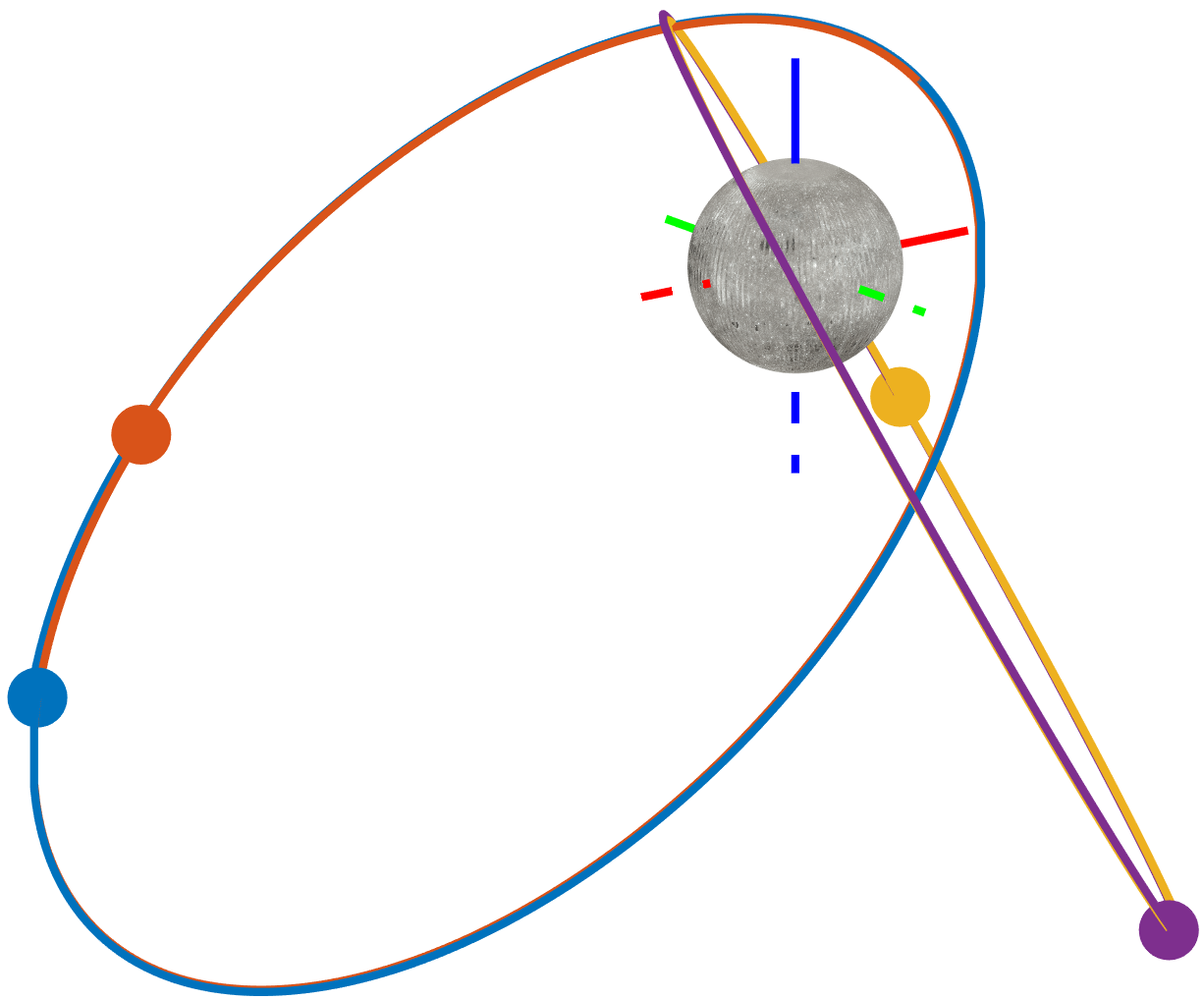}
		\caption{Orbits of four lunar satellites following \citet{audet2024} in Moon inertial frame with x-, y- and z-axis in red, green and blue}
		\label{fig:4sats_orbits}
	\end{minipage}
	\hfill
	\begin{minipage}{0.5\textwidth}
		\centering
		\includegraphics{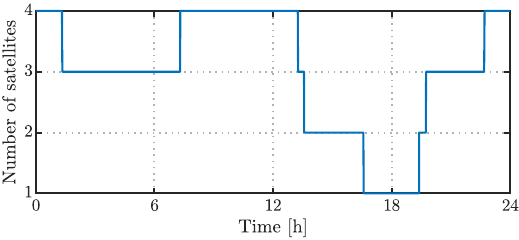}
		\caption{Number of visible satellites at landing site}
		\label{fig:4sats_visibility}
	\end{minipage}
\end{figure}
For the following case studies and simulations, we consider four lunar \ac{LCNS} satellites following \citet{audet2024}. Their orbits are shown in \cref{fig:4sats_orbits} in Moon inertial frame.
The satellites transmit LunaNet \ac{AFS} signals \citep{nasa2025a}, which are received by a satellite navigation receiver characterized by \citet{melman2022}. We follow the $C/N0$ calculation by \citet{melman2022}, including satellite and user antenna characteristics, which results in a $C/N0$ of \unit[31]{dB-Hz} to \unit[45]{dB-Hz}.
Pseudorange and pseudorange rate observation errors are defined by \cref{eq:prSatError,eq:prrSatError}, respectively,
with the respective temporally correlated bias model \cref{eq:satGMP} and
\ac{DLL} and \ac{FLL} noise variances are calculated following \citep{melman2022}.
For the cooperative navigation part, we consider the \ac{DLR} swarm communications and navigation system with the parameters stated in \cref{ss:coopPRError}. 
Cooperative pseudorange observation errors are defined by \cref{eq:prCoopError} with
temporally correlated \ac{GMP-1} pseudorange bias \cref{eq:coopGMP} and
variance from \cref{eq:tofCRB}, considering fading from the two-ray ground reflection model \cref{eq:Prx}.

We consider a landing site close to the lunar south pole with latitude $-89.45^\circ$ and longitude $222.69^\circ$. %
The number of visible satellites at the landing site is shown in \cref{fig:4sats_visibility}.
Our analysis focuses on robotic exploration. For short term, position and velocity of a robotic rover can be well predicted using a combination of wheel or visual odometry and inertial sensors. Thus, we assume a small velocity noise coefficient $\sigmaiv = \unit[0.001]{m/s^{1.5}}$. The robots move within \unit[1]{km} of the lading site with a speed of \unit[1]{m/s}.
For the hybrid navigation system of the lunar surface users, we consider space-grade \acp{OCXO} with clock noise coefficients $\sigmaic{1} = \unit[2.52 \cdot 10^{-23}]{s}$ and $\sigmaic{2} = \unit[3.03 \cdot 10^{-24}]{s^{-1}}$ \citep{axtal2023}.
For the lunar reference station, we assume a Rubidium clock with $\sigmaic{1} = \unit[1.22 \cdot 10^{-23}]{s}$ and $\sigmaic{2} = \unit[6.21 \cdot 10^{-28}]{s^{-1}}$.
We further consider a time interval of $T = \unit[1]{s}$ and
the prior uncertainties
$\sigma_{\bm{p}_i^0} = \unit[1000]{m}$,
$\sigma_{\bm{v}_i^0} = \unit[10]{m/s}$,
$\sigma_{\delta_i^0} = \unit[5]{us}$,
$\sigma_{\dot{\delta}_i^0} = \unit[100]{ppb}$.

In the following, we present four lunar navigation case studies using the \ac{BCRB} from \cref{s:bcrb}. The \ac{BCRB} provides a lower bound on the achievable position estimation \ac{MSE} and is therefore independent of choice and implementation of the actual navigation algorithm or filter.

\subsection{Importance of a temporally correlated Error Model}
\begin{figure}[htb]
	\begin{minipage}{0.48\textwidth}
		\centering
		\includegraphics{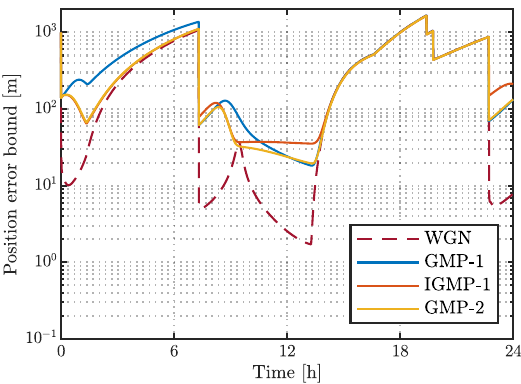}
		\caption{Satellite navigation position error bound of a single moving user using \ac{WGN}, \ac{GMP-1}, \ac{IGMP-1} and \ac{GMP-2} satellite navigation error models, respectively}
		\label{fig:4sats_1agent_GMP}
	\end{minipage}
	\hfill
	\begin{minipage}{0.48\textwidth}
		\centering
		\includegraphics{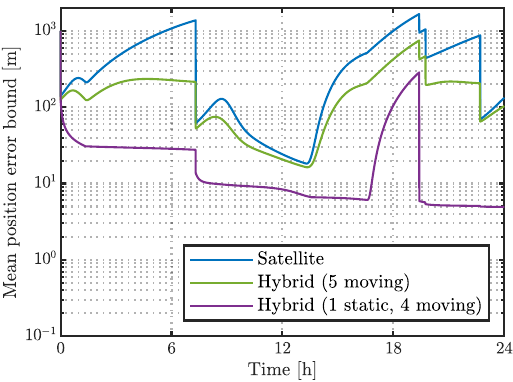}
		\caption{Mean position error bound of five lunar surface users considering satellite navigation and hybrid navigation with or without a static users, respectively}
		\label{fig:4sats_5agents_SatHybrid}
	\end{minipage}
\end{figure}

First, we compare the different satellite navigation error models introduced in \cref{s:satNav}.
\Cref{fig:4sats_1agent_GMP} shows the position error bound of a single moving user on the lunar surface using satellite navigation.
It is apparent that the zero mean \ac{WGN} model is often too optimistic.
In comparison, the resulting position error bounds using \ac{GMP-1}, \ac{IGMP-1} and \ac{GMP-2}, respectively, are fairly close together.
As expected, the position error bound using \ac{IGMP-1} grows over time. Thus, this model is suitable if ephemeris updates shall be modeled.
Finally, \ac{GMP-1} appears to be more conservative than \ac{GMP-2}.

The results underline the necessity of considering the temporal correlation of satellite navigation pseudorange and pseudorange rate observation errors.
First, the temporal correlation must be considered for studies to obtain realistic results for the expected position error.
Second, the temporal correlation must also be considered when implementing a navigation filter. Otherwise, the estimated covariance will be too optimistic, which can lead to filter divergence.
For the following studies and simulations, we model the satellite navigation errors by \ac{GMP-1}.

\subsection{Benefit of Hybrid Navigation}
Next, we examine the benefit of hybrid navigation, using the \ac{GMP-1} satellite navigation error model from \cref{ss:GMP1} and the cooperative navigation error model from \cref{ss:coopGMP}.
\Cref{fig:4sats_5agents_SatHybrid} shows the mean position error bound \cref{eq:peb} for five lunar surface users.
The baseline is satellite navigation without any cooperation among users.
By hybrid navigation, an accuracy gain can be achieved for the case where all five users are moving.
The accuracy gain is much more substantial when one user remains static. Such a static user can temporarily take the role of an anchor, albeit its position is not known perfectly. The static user considerably helps other users, who can benefit from accuracy gains by more than one order of magnitude. Furthermore, hybrid positioning is possible with only two visible satellites, whereas four are required for pure satellite navigation.
These results underline both, the navigation accuracy gain and the flexibility of hybrid navigation. The temporarily static user does not need any extra capabilities or a higher tier clock. It could e.g. be a robot within a team of robots, which keeps its position for a while to enable higher navigation accuracy for the other robots carrying out special tasks.

\subsection{Hybrid Navigation with Lunar Reference Station}
\begin{figure}[htb]
	\centering
	\includegraphics{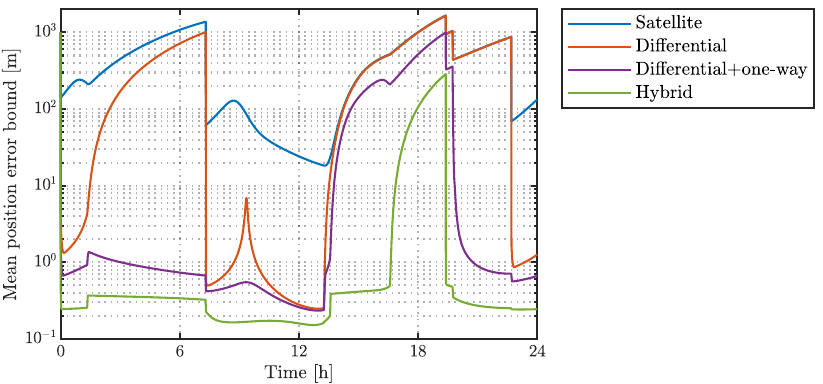}
	\caption{Mean position error bound of four moving lunar surface users using satellite navigation and three variants of a lunar reference station, namely differential navigation, differential navigation with one-way ranging of the reference station and hybrid navigation}
	\label{fig:4sats_1anchor_4agents_SatDiffBcHybrid_wide}
\end{figure}
Finally, we investigate the benefit of a lunar reference station with different capabilities. We assume the reference station has a Rubidium clock and its position on the lunar surface is precisely predetermined.
\Cref{fig:4sats_1anchor_4agents_SatDiffBcHybrid_wide} shows the mean position error bound \cref{eq:peb} of four moving surface users for different navigation modes.
The baseline is again satellite navigation, i.e. single point positioning with Doppler.
For differential navigation, we assume all users have pseudorange and pseudorange rate observations of the reference station available, which can be used to calculate a differential position solution. However, here we look at the \ac{BCRB} and are therefore independent of the actual algorithm used.
The differential navigation mode can improve position accuracy by up to two orders of magnitude - but only, when at least four satellites are visible. Furthermore, performance is highly dependent on the current satellite geometry as the error peak between \unit[8.5]{h} to \unit[10]{h} shows, which is attributed to high dilution of precision.
Additionally equipping the lunar reference station with one-way ranging capability decreases the minimum number of required satellites to three. Furthermore, the sensitivity w.r.t. to the current satellite geometry is reduced.
For hybrid navigation, satellite and cooperative navigation are combined, so users obtain pseudorange observations w.r.t. the reference station and vice versa.
Hybrid navigation provides the best accuracy in terms of the lowest position error bound of all investigated methods. The number of required satellites is further reduced to two.
Comparing hybrid navigation with a reference station to hybrid navigation with a temporarily static user analyzed earlier in \cref{fig:4sats_5agents_SatHybrid}, we see that the reference station provides a substantial accuracy gain. This is attributed to the precisely known position of the reference station and its Rubidium clock.

\section{Simulation Results}\label{s:results}
Based on the scenario defined in \cref{ss:scenario}, we have performed simulations to compare the performance of different hybrid navigation algorithms. For each setting, we have simulated 100 augmented state trajectories, see \cref{s:hybrid}. All filters experience the same noise realizations.
Baseline for comparison is the standard \ac{EKF}, where the satellite \ac{SISE} variance and the cooperative pseudorange variance are added to the observation variances, effectively ignoring temporal correlations of cooperative and satellite navigation errors.
We further take into account the augmented \ac{EKF}, augmented \ac{IEKF}, and augmented \ac{EKF-2} introduced in \cref{s:algorithms}, which fully consider temporal correlation of errors using the \ac{GMP-1} satellite navigation error model from \cref{ss:GMP1} and the cooperative navigation error model from \cref{ss:coopGMP}.
To ensure filter stability, updates are only performed when at least a total combined of three satellites and anchors are observed. 
Benchmark for all algorithms is the \ac{BCRB} from \cref{s:bcrb}, which is calculated for the simulated trajectories by numerically evaluating the expectation in \cref{eq:BIM}.

\begin{figure}[htb]
	\begin{minipage}{0.48\textwidth}
		\centering
		\includegraphics{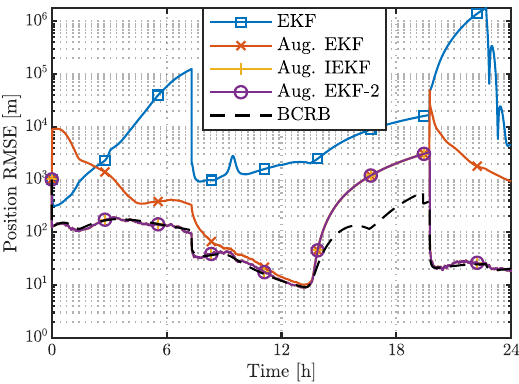}
		\caption{Hybrid navigation position RMSE of five moving lunar surface users for EKF, augmented EKF, augmented IEKF, EKF-2 and BCRB as benchmark}
		\label{fig:sim_4sats_5agents}
	\end{minipage}
	\hfill
	\begin{minipage}{0.48\textwidth}
		\centering
		\includegraphics{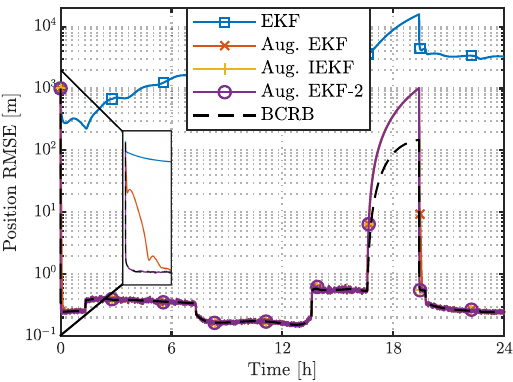}
		\caption{Hybrid navigation with lunar reference station position RMSE of four moving lunar surface users for EKF, augmented EKF, augmented IEKF, EKF-2 and BCRB as benchmark}
		\label{fig:sim_4sats_1anchor_4agents}
	\end{minipage}
\end{figure}
\cref{fig:sim_4sats_5agents} shows the position \ac{RMSE} of five moving lunar surface users.
The standard \ac{EKF} performs worst. Apparently, ignoring the error temporal correlation leads to an unstable filter.
The augmented \ac{EKF} appears to have convergence problems. Despite being widely used for satellite navigation, it cannot cope well with   stronger nonlinearities occurring in hybrid navigation.
The augmented \ac{IEKF} and augmented \ac{EKF-2} perform equally well. When at least three satellites are visible, their position \acp{RMSE} are close to the \ac{BCRB}. When less than three satellites are available, the position \acp{RMSE} deviate from the bound, as no filter updates are performed. If optimal performance in this challenging case is desired, one could consider e.g. batch algorithms.

Next, we investigate the scenario with a lunar reference station with precisely known position and Rubidium clock.
\Cref{fig:sim_4sats_1anchor_4agents} shows the position \ac{RMSE} of four moving surface users having connections to the reference station.
Again, the standard \ac{EKF}, which ignores the temporal correlation, is unstable.
In this case, the augmented \ac{EKF} often performs close to the \ac{BCRB} when at least two satellites plus the reference station are visible. However, it still has longer convergence times.
The augmented \ac{IEKF} and augmented \ac{EKF-2} show equal performance very close to the \ac{BCRB}. Due to lower complexity compared to augmented \ac{EKF-2}, the augmented \ac{IEKF} is the preferred algorithm for hybrid navigation, both with and without a reference station.

\begin{figure}[htb]
	\centering
	\includegraphics{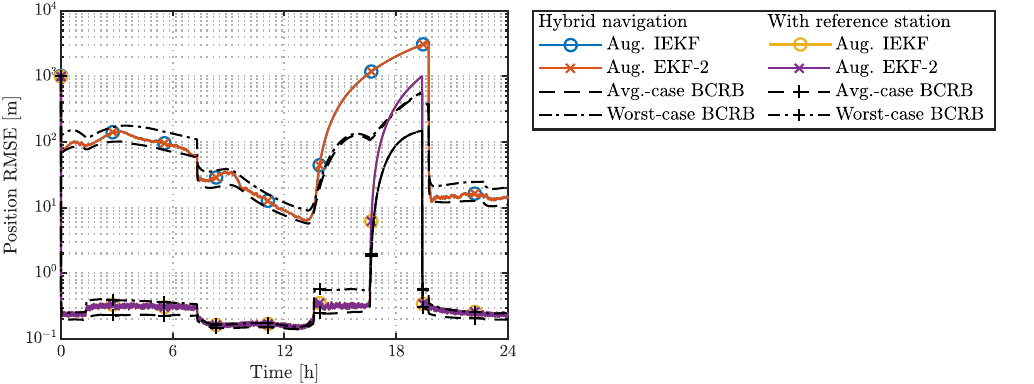}
	\caption{Position RMSE for augmented IEKF and augmented EKF-2 with worst-case parameters while actual error distributions follow average-case parameters and respective BCRBs}
	\label{fig:sim_simAvg_filterWorst}
\end{figure}

Finally, we investigate the practically relevant case where the filters consider worst-case \ac{GMP-1} parameters, while the the actual error distributions follow the average-case \ac{GMP-1} parameters. The parameters for satellite navigation are stated in \cref{ss:satParams} and for cooperative navigation in \cref{tab:CoopGMP}, respectively.
\Cref{fig:sim_simAvg_filterWorst} shows the respective \acp{RMSE} of augmented \ac{IEKF} and augmented \ac{EKF-2} for hybrid navigation with and without a lunar reference station, as well as the respective average-case and worst-case \acp{BCRB}.
For both cases, the filters using worst-case parameters slightly deviate from the average-case \ac{BCRB} matching the simulation settings. However, when enough satellites are visible, they still perform better than the worst-case \ac{BCRB}, which is shown for comparison.
Thus, using worst-case parameters for the filter, we can trade a slight loss of performance in the average-case for increased robustness w.r.t. the worst-case. 

\section{Conclusion}\label{s:conclusion}
In this paper, we have introduced and applied hybrid navigation error models for lunar surface users.
We have shown that the spatially correlated estimation bias due to multipath propagation has a decisive impact on the cooperative pseudorange error. Considering user movement, the cooperative pseudorange bias becomes temporally correlated and can be modeled by a \ac{GMP-1}.
We have also presented three different errors models for the lunar satellite navigation \ac{SISE} of pseudorange and pseudorange rate based on \ac{GMP-1},  \ac{IGMP-1}, and \ac{GMP-2}. 
Investigations of the hybrid navigation position error bound for a realistic scenario have revealed that the baseline \ac{WGN} model is often too optimistic, underlining the necessity to consider \ac{SISE} temporal correlation.
We have further shown a position accuracy gain by hybrid navigation compared to satellite navigation.
The accuracy gain is considerably higher when a static user is present, e.g. a temporarily standing robot. In this case, hybrid navigation with only two visible satellites is feasible.
We have also investigated the benefit of a lunar reference station with precisely known position and atomic clock.
Confirming results from literature, we have seen a substantial position accuracy gain by differential navigation, but only when at least four satellites are visible.
By transmitting a navigation signal, the reference station can replace one satellite.
The best operation mode for the reference station is hybrid navigation. In a network with four moving users, hybrid navigation achieves sub-meter position accuracy with only two visible satellites.

Finally, we have assessed different hybrid navigation filters by simulation.
The standard \ac{EKF}, which ignores the temporal error correlation, is not stable.
From the filters applying the introduced temporally correlated error models for cooperative and satellite navigation, the augmented \ac{EKF} has convergence problems.
Both, the augmented \ac{IEKF} and the augmented \ac{EKF-2} perform very close to the benchmark given by the \ac{BCRB}. The \ac{IEKF} is preferred due to lower complexity.
Using worst-case parameters for the filters, while the actual errors are distributed according to average-case parameters, incurs only a small performance penalty but makes the filters more robust.
In a nutshell, hybrid lunar \ac{PNT} is feasible and accurate, especially with a lunar reference station.

\section*{Acknowledgments}
The authors would like to thank Benjamin Siebler and Omar García Crespillo for fruitful discussions.

\bibliography{bibliography_abrv}
\bibliographystyle{apacite}

\end{document}

%% file: mydefs_arxiv.tex
\usepackage{layouts}

\usepackage{subcaption}

\usepackage{dsfont}
\usepackage{amsmath,amsfonts}
\usepackage{bm}
\usepackage{mathtools}
\usepackage{units}

\usepackage[nolist]{acronym}

\usepackage{booktabs}

\usepackage{todonotes}
\setuptodonotes{inline}

\usepackage[hidelinks]{hyperref}

\usepackage{doi}
\usepackage[natbibapa]{apacite} 

\usepackage[noabbrev,capitalise]{cleveref}
\crefname{appsec}{Appendix}{Appendices}
\crefname{equation}{}{}
\crefname{figure}{Fig.}{Figs.}

\newcommand{\cj}{\mathrm{j}}

\DeclarePairedDelimiter\abs{\lvert}{\rvert}%
\DeclarePairedDelimiter\norm{\lVert}{\rVert}%

\newcommand{\trace}[1]{\operatorname{tr}\left\{#1\right\}}
\newcommand{\diag}[1]{\operatorname{diag}\left\{#1\right\}}

\newcommand{\E}[1]{\operatorname{E}\left\{#1\right\}}
\newcommand{\var}[1]{\operatorname{var}\left\{#1\right\}}

\newcommand{\MSE}[1]{\operatorname{MSE}\left\{#1\right\}}

\newcommand{\EE}[2]{\operatorname{E}_{#1}\left\{#2\right\}}

\DeclareMathSymbol{\shortminus}{\mathbin}{AMSa}{"39}
\DeclareMathAlphabet{\mymathbb}{U}{BOONDOX-ds}{m}{n}

%% file: acronyms.tex
\acro{FOV}{field of view}
\acro{5G}{fifth generation}
\acro{6G}{sixth generation}
\acro{UAV}{unmanned aerial vehicle}
\acro{WSN}{wireless sensor networks}
\acro{RADAR}{radio detection and ranging}
\acro{IoT}{Internet of Things}
\acro{UWB}{ultra-wideband}
\acro{RF}{radio frequency}
\acro{IEEE}{Institute of Electrical and Electronics Engineers}

\acro{EM}{electromagnetic}
\acro{EADF}{effective aperture distribution function}
\acro{MMA}{multi-mode antenna}
\acro{TCM}{theory of characteristic modes}
\acro{ULA}{uniform linear array}
\acro{UCA}{uniform circular array}
\acro{RHCP}{right-hand circular polarization}
\acro{LHCP}{left-hand circular polarization}
\acro{AIT}{array interpolation technique}
\acro{WM}{wavefield modeling}
\acro{URA}{uniform rectangular array}
\acro{DoA}{direction-of-arrival}
\acrodefplural{DoA}[DoAs]{directions-of-arrival}
\acro{DoD}{direction-of-departure}
\acrodefplural{DoD}[DoAs]{directions-of-departure}
\acro{LOFAR}{low frequency antenna array}

\acro{LoS}{line-of-sight}
\acro{NLoS}{non-line-of-sight}
\acro{FSPL}{free-space path loss}
\acro{SNR}{signal-to-noise ratio}
\acro{MIMO}{multiple-input multiple-output}
\acro{TDMA}{time-division multiple access}
\acro{SOTDMA}{self-organized time-division multiple access}
\acro{PSD}{power spectral density}
\acro{RTD}{round-trip delay}
\acro{RTT}{round-trip time}
\acro{TWR}{two-way ranging}
\acro{RSS}{received signal strength}
\acro{OFDM}{orthogonal frequency-division multiplexing}
\acro{ToA}{time-of-arrival}
\acro{ToT}{time-of-transmission}
\acro{ToF}{time-of-flight}
\acro{CW}{continuous wave}
\acro{CFO}{carrier frequency offset}
\acro{PSK}{phase-shift keying}
\acro{RFID}{radio-frequency identification}
\acro{GT}{guard time}
\acro{CP}{cyclic prefix}
\acro{mmWave}{millimeter wave}
\acro{QPSK}{quadrature phase-shift keying}
\acro{CDMA}{code division multiple access}

\acro{CDF}{cumulative distribution function}
\acro{MAP}{maximum a posteriori}
\acro{CRB}{Cram\'{e}r-Rao bound}
\acro{BCRB}{Bayesian Cram\'{e}r-Rao bound}
\acro{KLF}{Kalman Laplace filter}
\acro{IKF}{iterated Kalman filter}
\acro{MSE}{mean squared error}
\acro{FFT}{fast Fourier transform}
\acro{EKF}{extended Kalman filter}
\acro{IEKF}{iterated extended Kalman filter}
\acro{EKF-2}{second order extended Kalman filter}
\acro{pdf}{probability density function}
\acro{ML}{maximum likelihood}
\acro{FIM}{Fisher information matrix}
\acro{EFIM}{equivalent Fisher information matrix}
\acro{BIM}{Bayesian information matrix}
\acro{EBIM}{equivalent Bayesian information matrix}
\acro{RMSE}{root-mean-square error}
\acro{i.i.d.}{independent and identically distributed}
\acro{PEB}{position error bound}
\acro{OEB}{orientation error bound}
\acro{AIC}{Akaike information criterion}
\acro{BIC}{Bayesian information criterion}
\acro{GLRT}{generalized likelihood ratio test}
\acro{ZZB}{Ziv-Zakai bound}
\acro{WWB}{Weiss-Weinstein bound}
\acro{ICOMP}{information complexity criterion}
\acro{ACF}{autocorrelation function}

\acro{MUSIC}{multiple signal characterization}
\acro{BSS}{blind source separation}
\acro{ESPRIT}{estimation of signal parameters by rotational invariance techniques}
\acro{IQML}{iterative quadratic maximum likelihood}
\acro{SAGE}{space-alternating generalized expectation maximization}
\acro{BFGS}{Broyden–Fletcher–Goldfarb–Shanno}
\acro{DFT}{discrete Fourier transform}
\acro{SLAM}{simultaneous localization and mapping}
\acro{SLAC}{simultaneous localization and calibration}
\acro{MVDR}{minimum variance distortionless response}

\acro{MEMS}{microelectromechanical system}
\acro{IMU}{inertial measurement unit}
\acro{USRP}{Universal Software Radio Peripheral}
\acro{SDR}{software-defined radio}
\acro{RTK}{real-time kinematic}
\acro{LO}{local oscillator}
\acro{HIL}{hardware-in-the-loop}
\acro{COTS}{commercial off-the-shelf}
\acro{RFIC}{radio-frequency integrated circuit}
\acro{FPGA}{field programmable gate array}

\acro{RC}{reduced complexity}
\acro{FoV}{field of view}
\acro{w.r.t.}{with respect to}
\acro{DLR}{German Aerospace Center}
\acro{NASA}{National Aeronautics and Space Administration}
\acro{LRU}{leightweight rover unit}
\acro{ISECG}{International Space Exploration Coordination Group}
\acro{TRL}{technology readiness level}
\acro{DSN}{deep space network}

\acro{ECEF}{Earth-centered, Earth-fixed}
\acro{PA}{Principal Axis}
\acro{GPS}{Global Positioning System}
\acro{GNSS}{Global Navigation Satellite System}
\acro{PNT}{positioning, navigation and timing}
\acro{OCXO}{oven-controlled crystal oscillator}
\acro{DLL}{delay-locked loop}
\acro{FLL}{frequency-locked loop}
\acro{PLL}{phase-locked loop}
\acro{GMAT}{General Mission Analysis Tool}
\acro{DEM}{digital elevation model}
\acro{ELFO}{elliptical Lunar frozen orbit}
\acro{SISE}{signal-in-space error}
\acro{IGS}{International GNSS Service}
\acro{PRN}{pseudo-random noise}
\acro{PR}{pseudorange}
\acro{PRR}{pseudorange rate}

\acro{LNSS}{Lunar Navigation Satellite System}
\acro{LCNS}{Lunar Communication and Navigation Services}
\acro{LCRNS}{Lunar Communication Relay and Navigation System}
\acro{LNIS}{LunaNet Interoperability Specification Docucment}
\acro{AFS}{augmented forward signal}
\acro{LuGRE}{Lunar GNSS Receiver Experiment}

\acro{TTL}{time to live}
\acro{ERV}{equivalent ranging variance}

\acro{WGN}{white Gaussian noise}
\acro{GMP-1}{first order Gauss-Markov process}
\acro{IGMP-1}{integrated first order Gauss-Markov process}
\acro{GMP-2}{second order Gauss-Markov process}